\documentclass[twocolumn]{aastex631}

\usepackage{enumitem}
\usepackage{amsmath}
\usepackage{latexsym}
\usepackage{graphicx}
\usepackage{epsf}
\usepackage{CJK}
\usepackage{wasysym}
\usepackage{natbib}

\newcommand{\oiii}{[O\,{\sc iii}]}
\newcommand{\nii}{[N\,{\sc ii}]}
\newcommand{\civ}{C\,{\sc iv}}
\newcommand{\ha}{H$\alpha$}
\newcommand{\hb}{H$\beta$}
\newcommand{\feii}{Fe\,{\sc ii}}

\newcommand{\nh}{N_\mathrm{H}}
\newcommand{\cm}{\mathrm{cm}}
\newcommand{\kev}{\,keV}
\newcommand{\WISE}{\textit{WISE}}

\received{December 8th, 2023}
\revised{August 4th, 2024}
\accepted{August 6th, 2024}

\submitjournal{ApJ}

\shorttitle{Extremely Red Quasars in X-Rays}
\shortauthors{Ma et al.}

\graphicspath{{./}{figures/}}

\begin{document}
\begin{CJK*}{UTF8}{gbsn}
\title{Evidence for Intrinsic X-ray Weakness Among Red Quasars at Cosmic Noon}

\correspondingauthor{Yilun Ma}
\author[0000-0002-0463-9528]{Yilun Ma (马逸伦)}
\affiliation{Department of Astrophysical Sciences, Princeton University, Princeton, NJ 08544, USA}
\email{yilun@princeton.edu}

\author[0000-0003-4700-663X]{Andy Goulding}
\affiliation{Department of Astrophysical Sciences, Princeton University, Princeton, NJ 08544, USA}

\author[0000-0002-5612-3427]{Jenny E. Greene}
\affiliation{Department of Astrophysical Sciences, Princeton University, Princeton, NJ 08544, USA}

\author[0000-0001-6100-6869]{Nadia L. Zakamska}
\affiliation{Department of Physics \& Astronomy, Johns Hopkins University, Baltimore, MD 21218, USA}

\author[0000-0003-2212-6045]{Dominika Wylezalek}
\affiliation{Zentrum f\"ur Astronomie der Universit\"at Heidelberg, Astronomisches Rechen-Institut, \\M\"onchhofstra{\ss}e 12-14, D-69120 Heidelberg, Germany}

\author[0000-0002-2624-3399]{Yan-Fei Jiang (姜燕飞)}
\affiliation{Center for Computational Astrophysics, Flatiron Institute, New York, NY 10010, USA}

\begin{abstract}
Quasar feedback is a key ingredient in shaping galaxy evolution. A rare population of extremely red quasars (ERQs) at $z=2-3$ are often associated with high-velocity \oiii$\lambda5008$ outflows and may represent sites of strong feedback. In this paper, we present an X-ray study of 50 ERQs to investigate the link between the X-ray and outflow properties of these intriguing objects. Using hardness ratio analysis, we confirm that the ERQs are heavily obscured systems with gas column density reaching $N_\mathrm{H}=10^{23-24}\,\mathrm{cm^{-2}}$. We identify 20 X-ray-non-detected ERQs at high mid-infrared luminosities of $\nu L_\mathrm{\nu,6\mu m}\gtrsim3\times10^{46}\,\mathrm{erg\,s^{-1}}$. By stacking the X-ray observations, we find that the non-detected ERQs are on average underluminous in X-rays by a factor of \textcolor{black}{$\sim10$} for their \textcolor{black}{mid-infrared} luminosities. We consider such X-ray weakness to be due to both heavy gas absorption and intrinsic factors. Moreover, we find that the X-ray-weak sources also display higher-velocity outflows. One option to explain this trend is that weaker X-rays facilitate more vigorous line-driven winds, which then accelerate the \oiii-emitting gas to kpc-scales. Alternatively, super-Eddington accretion could also lead to intrinsic X-ray weakness and more powerful continuum-driven outflow. 

\end{abstract}


\section{Introduction}\label{sec:intro}
It has been long thought that the supermassive black holes (SMBHs) at the center of galaxies could significantly impact and coevolve with their hosts via feedback processes \citep[e.g.,][]{King2003, Veilleux2005, HeckmanBest2014}. These feedback processes may be responsible for the high-end cutoff of the galaxy luminosity function \citep{Croton2006} and the required rapid quenching mechanisms to produce the observed massive quiescent galaxy populations at high redshift \citep[e.g.,][]{Girelli2019, Ito2023}. For example, accreting SMBHs can become luminous enough to launch galactic outflows that are capable of removing gas and therefore quenching the star formation of the host galaxy \citep{SilkRees1998, Hopkins2006}. This is often referred to as ``quasar-mode" feedback. Observationally, the presence of quasar-driven outflows and winds is often inferred from blueshifted and/or asymmetric emission or absorption line profiles in both spatially unresolved and resolved spectroscopic observations \citep[e.g.,][]{Cano-Diaz2012, Greene2012, Harrison2016, ForsterSchreiber2019, Leung2019}.

Both the cosmic star formation rate density and quasar activities peak at $z=2-3$, a period also known as ``Cosmic Noon" \citep{BoyleTerlevich1998, MadauDickinson2014}. Feedback is thus believed to be the strongest during this period. Indeed, in this redshift range, a population of extremely red quasars (ERQs) is identified based on their high infrared-to-optical ratios and high rest equivalent width of ultraviolet emission lines \citep{Ross2015, Hamann2017}. These ERQs are often found to host extreme ionized \oiii$\lambda5008$ outflows up to $|\Delta v| \approx 1500\,\mathrm{km\,s^{-1}}$, unmatched by any other quasar populations \citep{Zakamska2016, Perrotta2019}. Thus, despite their rarity \citep{Hamann2017},  ERQs represent the perfect sites to study strong feedback at the peak of quasar activity in cosmic time. Consequently, extensive multiwavelength observations have been carried out for the intriguing ERQ population across the electromagnetic spectrum \citep[e.g.,][]{Alexandroff2018, Hwang2018, Vayner2021, Ishikawa2021, Vayner2023}. Moreover, recent \textit{James Webb Space Telescope} ({\it JWST}) observation reveals that at least some ERQs are signposts of protocluster formation \citep{Wylezalek2022}, leaving the environmental effects on this quasar population an open front of investigation. 

Based on their unique colors and extreme outflows/winds at different scales, the ERQs may also be the rare examples of luminous super-Eddington accretors at Cosmic Noon (see \citealt{Alexandroff2018}), but no consensus on the black hole masses of the ERQs has been reached yet \citep{Perrotta2019, Zakamska2019}. Moreover, another challenge is to determine the bolometric luminosity of the ERQs given their obscured nature. X-ray observations therefore become particularly useful to gauge the level of circumnuclear obscuration and estimate the intrinsic luminosities of these targets. \textcolor{black}{\cite{Goulding2018} conducted a pilot X-ray study on 11 ERQs (10 at $z>2$ and 1 at $z\sim1.5$) with confirmed \oiii{} outflows and/or usable observations from \textit{XMM-Newton} and \textit{Chandra X-ray Observatory} at the time.} All 11 ERQs yield significant detections, and both X-ray spectroscopy and hardness ratio analysis suggest that all are heavily obscured sources with $N_\mathrm{H}\approx10^{23}\,\mathrm{cm^{-2}}$. Having been corrected for gas absorption, these ERQs appear to be intrinsically as luminous in X-ray as the unobscured quasars at similar \textcolor{black}{mid-infrared} luminosities. \textcolor{black}{Intriguingly, given that quasars with strong radiatively-driven outflows tend to be weak in X-ray \citep[e.g.,][]{Luo2014, Ricci2017}, two questions then emerge from these results: 1) what is the driving mechanisms of the extreme outflows in ERQs? 2) does there exist a population of X-ray weak ERQs?}



In this work, we present our analysis of a more extensive X-ray sample of 40 ERQs in addition to the \cite{Goulding2018} targets in an attempt to answer the questions above. \S\,\ref{sec:sample_obs_reduction} introduces our samples, X-ray observation with \textit{Chandra}, ground-based spectroscopic observations in the rest-frame optical, and their respective data reduction procedures. \S\,\ref{sec:analysis} specifies our procedures for spectroscopic and X-ray analysis. \S\,\ref{sec:result} lists our results. We then discuss the results and their further implications in \S\,\ref{sec:discussion} and conclude in \S\,\ref{sec:conclusion}. Throughout the paper, we assume a $H_0=70\,\mathrm{km\,s^{-1}\,Mpc^{-1}}$, $\Omega_m = 0.3$, and $\Omega_\Lambda=0.7$ cosmology. The emission lines are identified by their vacuum wavelengths. 

\section{Sample, Observation, and Data Reduction}\label{sec:sample_obs_reduction}
\subsection{ERQ Parent Sample}
\begin{figure}
    \centering
    \includegraphics[width=\columnwidth]{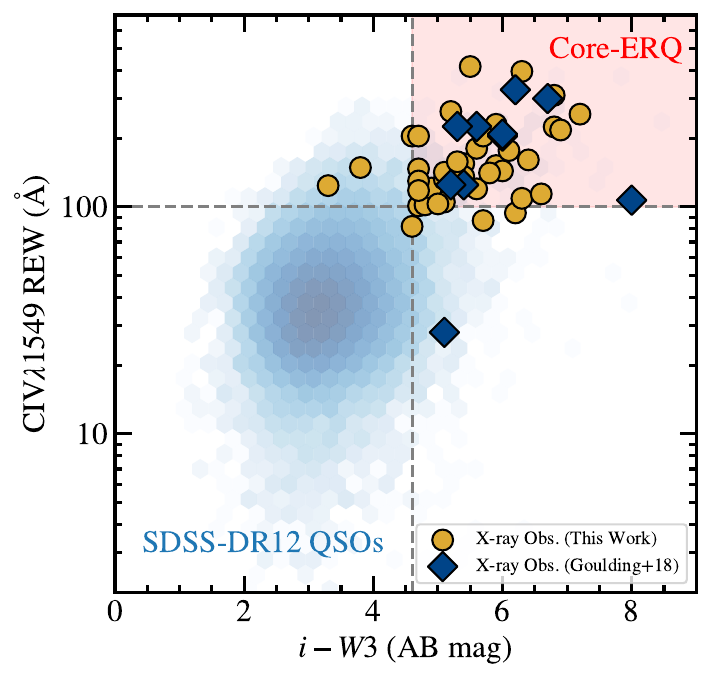}
    \caption{The selection criteria of the ERQ population. The blue hexagons are adopted from the SDSS DR12 QSO catalog.}
    \label{fig:erq_selection} 
\end{figure}

A population of 65 ERQs was first identified by \cite{Ross2015} for their high infrared-to-optical flux ratio ($r_\mathrm{AB}-W4_\mathrm{Vega}>14\,\mathrm{mag}$) by cross-matching quasar catalogues of Sloan Digital Sky Survey (SDSS) and Baryonic Oscillation Spectroscopic Survey (BOSS; \citealt{Dawson2013}, \citealt{Paris2014}) to that of the \textit{Wide-Field Infrared Survey Explorer} (\WISE). The redshift of these quasars ranges over $0.3\lesssim z \lesssim 4.4$, peaking at $z\sim0.8$ and $z\sim 2.5$. In the UV wavelengths, the population shows many intriguing spectroscopic properties such as wingless \civ$\lambda1549$ profiles, peculiar emission line ratios, and features of both type-1 (unobscured; e.g., broad permitted lines) and type-2 (obscured; e.g., continuum suppression) quasars. \textcolor{black}{It is therefore not possible to place the whole ERQ population into either type-1 or type-2 quasars, although the presence of broad lines often implies that this intermediate population is more type-1-like.}

Using the BOSS quasar catalogue for Data Release 12 (DR12Q; \citealt{Paris2014, Paris2017}), \cite{Hamann2017} refined the definition of ERQs based on color and rest equivalent width (REW) of the \civ$\lambda1549$ lines. Primarily, ERQs are selected by their red colors, i.e. $i-W3\geqslant4.6\,\mathrm{mag}$ (AB magnitude). \textcolor{black}{Furthermore, a core subsample of 97 ERQs, dubbed``core-ERQs", is identified by REW(\civ)$\,\geqslant100\,\mathrm{\AA}$ at $2<z<3.4$ by the similarity of their distinctive emission-line properties and shapes of the spectral energy distribution (SED).} We show the selection of core-ERQs and their comparison with the SDSS-DR12 QSOs at $2<z<3$ in Figure \ref{fig:erq_selection}. \cite{Hamann2017} also expand the samples to include 235 ``ERQ-like" quasars that meet either of the two ``core" criteria mentioned above or whose \civ{} profile is wingless/boxy inferred from the line profile's kurtosis measurements. For the rest of the paper, we make no distinction between core and ERQ-like targets and dub both sub-populations ``ERQs", since the majority of our sample is drawn from the core population and only \textcolor{black}{six} sources are significantly outside of the core-ERQ selection box (see Figure \ref{fig:erq_selection} and \S\,\ref{sec:xray_sample} below). 

\subsection{ERQ X-Ray Sample and Observation}\label{sec:xray_sample}
\begin{figure}
    \centering
    \includegraphics[width=\columnwidth]{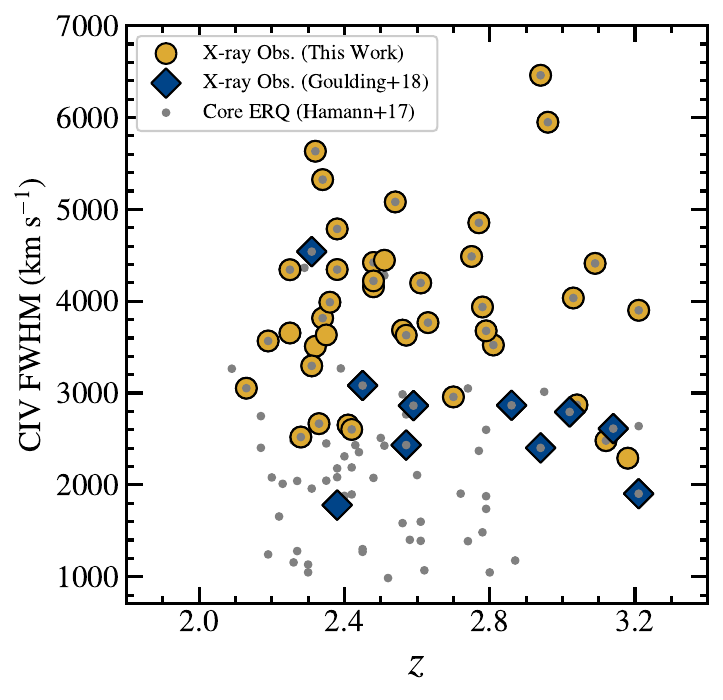}
    \caption{\textcolor{black}{The sample of ERQ targets with X-ray observations is complete for FWHM(\civ)$\,>3500\,\mathrm{km\,s^{-1}}$. The gray dots are the core-ERQ sources from \cite{Hamann2017}. Note that the blue diamond at $z\approx3.2$ is SDSS J212951.40$-$001804.3: \cite{Hamann2017} classified it as ERQ-like in its original catalog due to non-detection in W3; however, \cite{Goulding2018} point out that this source fulfills the core-ERQ the requirement if one uses the most up-to-date \textit{WISE} photometry.}}
    \label{fig:fwhm_redshift}
\end{figure}
\textcolor{black}{In order to measure the full SED and potentially estimate bolometric luminosity, probe circumnuclear obscuration level of the ERQs,} and to link the X-ray properties to the high-velocity outflows in these systems, we proposed \textit{Chandra} observations (PI: A. Goulding) in Cycle 22 for 23 ERQ targets with FWHM(\civ)$\,\gtrsim3500\,\mathrm{km\,s^{-1}}$ \textcolor{black}{and no scientifically usable archival X-ray data at the time.} The selection using \civ{} instead of \oiii{} is due to the lack of rest-frame optical spectra for all the targets. However, the asymmetric profile and the broad line width of the \civ$\lambda1549$ line hint at the presence of winds \citep{Hamann2017, Alexandroff2018}, so \civ{} can possibly be used to estimate the velocity of \oiii{}-emitting outflow (see \S\,\ref{sec:erq_line_properties}) if the two gases are coupled. All targets were approved and observed in the observing cycle. We also cross-matched the \textit{Chandra} archival data against the ERQ catalogue of \cite{Hamann2017} and include 16 more \textit{Chandra} targets on the ACIS-S S3 chip or the ACIS-I chips with mostly snapshot observations ($\sim4\,\mathrm{ks}$) in addition to our deeper ($\sim10-20\,\mathrm{ks}$) Cycle 22 campaign.  

In total, we build an X-ray sample of 40 ERQ targets for this work. We show in Figure~\ref{fig:fwhm_redshift} that our sample is complete to ERQs with FWHM(\civ)$\,\gtrsim3500\,\mathrm{km\,s^{-1}}$. In addition, jointly with 10 ERQs at $z > 2$ studied by \cite{Goulding2018}, we probe the X-ray properties of ERQs on the higher end of their outflow velocities. \textcolor{black}{We list both the new X-ray sample (40 ERQs) and the 10 ERQs from \cite{Goulding2018} in Table~\ref{tab:sample}. }

For consistency with the analyses of ERQs presented in \cite{Goulding2018}, we follow the same processing and data reduction used previously. Briefly, data processing is carried out using the standard Chandra X-ray Center software packages in {\tt CIAO} v4.15 \citep{ciao_citation}. Consistent calibrations from the CALDB 4.10.2 library are applied using {\tt chandra\_repro}. Streak events, bad pixels, pixel randomization and cosmic rays are removed with STATUS$=$0 and screened with the typical grad set during the implementation of {\tt acis\_process\_events}. Flares more than 3$\sigma$ above the background are identified and removed using the {\tt lc\_clean} package before creating final Level-2 events files. Final exposure times in Good Time Intervals are in the range $\sim$4--40~ks. We present the analysis of the reduced \textit{Chandra} X-ray data in \S\,\ref{sec:HR_LX_measurement}.

\subsection{Ground-Based Near-Infrared Spectroscopy of ERQs}
As follow-up observations, we obtained rest-frame optical spectra of 7 ERQs (SDSS J082649.30+163945.2, SDSS J085451.11+173009.1, SDSS J102130.74+214438.4, SDSS J110202.68$-$000752.7, SDSS J145623.35+214516.2, SDSS J150117.07+231730.9, and SDSS J154243.87+102001.5) among the X-ray sample using the Folded port InfraRed Echellette (FIRE) spectrograph on the Magellan Baade Telescope, which covers the wavelength range between 0.82 and 2.51$\,\mathrm{\mu m}$. At the redshift of these ERQs, the \oiii$\lambda\lambda4960,5008$ doublet and \hb{} emission lines fall into the observed \textit{H}- or \textit{K}-bands. Depending on the specific redshift of each individual source, we are also able to probe the H$\gamma$ and/or H$\alpha$ emission of some ERQs. For each target, individual exposures range from $450\,\mathrm{s}$ to $1200\,\mathrm{s}$, and the total integration times range from $0.75\,\mathrm{h}$ to $1.6\,\mathrm{h}$. We observed all targets with a slit width of 0.75 arcseconds. 

We reduce the FIRE spectra using the reduction pipeline {\tt\string PypeIt} \citep{pypeit:joss_arXiv, pypeit:zenodo}. The prominent OH emission lines in the spectra are used to calibrate the wavelength solution for each frame. For sources with a bright continuum, we perform automatic extraction in the pipeline. For those with weak continuum, we conduct a semi-automatic routine by manually identifying emission lines in the 2D spectra and therefore guide the code to extract at certain positions along each order of the echellete. The telluric correction and flux calibration are performed using the standard/telluric star spectra observed on the same nights. 

Since not all ERQs in the X-ray sample have readily available spectra, we also invoke the rest-frame optical spectra for 28 ERQs taken by \cite{Perrotta2019} in addition to our FIRE observations. This makes a spectroscopic sample of 35 ERQs with available rest-frame optical spectra in total. 19 of the 40 ERQs in our X-ray sample overlap with the spectroscopic sample, and 5 of the 10 ERQs in the X-ray pilot study by \cite{Goulding2018} overlap with this spectroscopic sample. 

\subsection{Additional HotDOG X-ray Sample as Comparison}\label{sec:hotdog_as_comparison_sample}
Hot dust-obscured galaxies (HotDOGs) are selected to be \textit{WISE} sources with no detection or marginal detections at 3.4 and 4.6$\,\mathrm{\mu m}$ but strong detections at 12 and 22 $\,\mathrm{\mu m}$ likely produced by dust illuminated by the central SMBHs \citep{Eisenhardt2012, Wu2012, Bridge2013}. The SEDs of HotDOGs are very similar to those of ERQs, characterized by a relatively flat slope in the UV and a steep rise in the between $\sim1$ to 3 microns, but the UV-to-optical continuum is more suppressed and the near-to-mid-infrared rise is steeper in HotDOGs than in the ERQs \citep{Tsai2015, Hamann2017}. Although selected using different criteria, HotDOGs and ERQs may both represent a similar phase of obscured accretion, with the former being more dust-dominated. Therefore, we include an X-ray sample of 18 HotDOGs from \cite{Vito2018} in our study. This allows us to probe and compare the X-ray properties of red quasars at even \textcolor{black}{higher luminosities ($\nu L^\mathrm{HotDOG}_\mathrm{\nu,6\,\mu m}\approx 10^{47-48}\,\mathrm{erg\,s^{-1}}$) than the ERQs ($\nu L^\mathrm{ERQ}_\mathrm{\nu,6\,\mu m}\approx 10^{46-47}\,\mathrm{erg\,s^{-1}}$)}. We are not supplemented with rest-frame optical spectra for the HotDOGs; as a population, they do not show as powerful outflows as the ERQs do \citep{Wu2018, Finnerty2020}. Conducting a detailed analysis of their outflows is beyond the scope of this work. 

\section{Analysis}\label{sec:analysis}


\subsection{Emission Line Analysis on Rest-Frame Optical and UV Spectra}\label{sec:line_fitting}
ERQs show signatures of the most extreme outflows in the universe. Their \oiii{} emission is so broadened and blueshifted that it is strongly blended with the H$\beta$ emission \citep{Zakamska2016, Perrotta2019}. Spectroscopic analysis of ERQs is essential to determine their redshifts and infer the gas kinematics in these sources. Therefore, we conduct multi-component fitting to the ERQs' rest-frame optical and UV spectra. 

For ERQs with rest-frame optical spectra available (from either our FIRE observations or \citealt{Perrotta2019}), we fit the rest-frame optical spectra over the entire observed \textit{H}- and \textit{K}- bands (if available). We characterize the continuum as two independent linear functions in the two bands, respectively. We model the emission line profiles as Gaussians. For each line in the rest-frame optical spectra, we allow there to be a maximum of three kinematic components: 
\begin{enumerate}
    \item a broad component for the permitted lines with full-width-half-maximum (FWHM) $>2000\,\mathrm{km\,s^{-1}}$ \citep{Zakamska2003} to represent the broad-line region (BLR) emission;
    \item a narrow component for both permitted and forbidden lines with FWHM $<2000\,\mathrm{km\,s^{-1}}$ centered with the BLR component to represent the narrow-line region (NLR) emission -- all narrow lines have the same FWHM;
    \item a blueshifted component for both permitted and forbidden lines representing any additional velocity component such as outflows. 
\end{enumerate}
\textcolor{black}{Within each of the three categories above, we fix the FWHM of all the lines under the assumption that all line emission originates from gas with the same kinematics. For example, if we have H$\beta$ emission with all three components and an \oiii{} emission with only the NLR and outflow components, the fitting procedure requires $\mathrm{FWHM_{BLR}^{H\beta}} \neq \mathrm{FWHM_{NLR}^{H\beta}} =  \mathrm{FWHM_{NLR}^{[OIII]}}$ and $\mathrm{FWHM_{outflow}^{H\beta}} = \mathrm{FWHM_{outflow}^{[OIII]}}$.} 
Limited by the signal-to-noise ratio (SNR) in some of the spectra, we also fix the NLR-to-outflow flux ratio across all the lines. This treatment aims to reduce the fitting degeneracy due to the blending of \hb+\oiii{} (and sometimes \ha+\nii{}) complexes. It has been discussed in the literature that \oiii{} and \hb{} could be kinematically untied particularly in the strong-outflow regime \citep{Zakamska&Greene2014, Zakamska2016} due to processes such as NLR stratification \citep{Osterbrock1991, Komossa2018} and therefore a more agnostic procedure may be more suitable for fitting these ERQs. For example, \cite{Perrotta2019} assign no physical meanings (i.e. NLR or BLR) to the Gaussian components in their fits and allow different kinematics for \oiii{} and \hb{}. However, we find that our NLR/BLR treatment also provides equally reasonable residuals and untying the \oiii{} and \hb{} kinematics largely yields no significant improvements for most of our ERQs. 

We let the FWHM of the NLR emission be larger than $2000\,\mathrm{km\,s^{-1}}$ in a few ERQ targets in order to capture the full profiles. Based on recent {\it JWST} observation, the immediate environment of ERQs may be complex and dynamic with companion galaxies and gas clumps in addition to the quasar outflow \citep{Wylezalek2022}. These velocity components may blend and result in line widths larger than those found in typical NLR emissions in our spatially unresolved spectra. 

We also allow the \feii{} emissions to exist if a BLR component for the Balmer lines is detected; conversely, if \feii{} features are identifiable by visual inspection, we consequently add a BLR component for the permitted lines. The \feii{} emission is modelled by convolving a Gaussian kernel that has the same FWHM in velocity space with the \cite{Veron-Cetty2004} \feii{} template. The continuum, the emission lines, and the \feii{} template are fitted simultaneously. 

There are sometimes ambiguities in choosing either a single BLR component for \hb{} or a wide blueshifted component for the \oiii{} to account for the excess flux between the emission lines. In those cases, we turn to the BLR-\hb{} scenario if the blue wing of the fitted \oiii$\lambda5008$ profile never dominates over any of the other kinematics components at any wavelengths. Such ambiguity is accounted for in the systematic errors in FWHM measurements. We show some examples of our fitting for the rest-frame optical spectra in Figure~\ref{fig:spectra}. 

We also fit the \civ$\lambda1549$ emission line of ERQs in both the X-ray and spectroscopic samples using their SDSS spectra. For ERQs with rest-frame optical spectra available, we fit the \civ{} line with the same number of components as \hb; otherwise, we fit the emission line with a maximum of three components as described above. \civ{} lines show many intriguing properties and hint at the obscured nature of ERQs \citep{Hamann2017}, but their origins in the ERQs remain poorly understood. 
A UV-emitting disk wind is a \textcolor{black}{possible} origin given the \civ{} is often blueshifted, but most such systems show REW(\civ) \textcolor{black}{much smaller than that of the ERQs} \citep{Richards2011}. \cite{ZakamskaAlexandroff2023} also argue that the \civ{} lines are likely significantly reprocessed and scattered by the torus skin wind, so they may originate on the intermediate spatial scales between those characteristic of the classical BLR and NLR. \textcolor{black}{Although many lines of evidence have been invoked, the spatial extent and origin of \civ{} remains ambiguous. }Thus, we do not assign physical meanings to any of these components -- the goal is to fully capture the shape of the \civ{} lines. Again, the emission feature and a linear continuum are fitted simultaneously within $\sim 500\,\mathrm{\AA}$ of the \civ{} line with any absorption features manually masked out. We also show examples of the \civ{} fitting in Figure~\ref{fig:spectra}. 

To determine the systemic redshift of the ERQs, we use the centroid of the NLR component of the luminous \oiii$\lambda5008$ line that is most likely associated with the host galaxy. 
For ERQs without a rest-frame optical spectrum, we determine the systemic redshift using the peak of the full \civ{} profile as our best available estimation. The resonant \civ{} line could be blueshifted as much as $\sim500\,\mathrm{km\,s^{-1}}$ from the true systemic redshift due to potential line-driven winds in the BLR (see \citealt{Richards2011}, \citealt{ZakamskaAlexandroff2023}, and references therein). \cite{Gillette2023} find that ERQs tend to show an even higher incidence of blueshifted \civ{} than blue quasars. However, the result of our study does not significantly depend on the accuracy of redshift measurement. 

\subsection{X-ray Analysis}\label{sec:HR_LX_measurement}
To analyze the X-ray data, we bin the \textit{Chandra} images into three observed energy bands: 0.3--1\,keV (soft), 1--4\,keV (middle), and 4-7\,keV (hard), which roughly correspond to 1--3.5\,keV, 3.5--14\,keV, and 14--24.5\,keV in the rest frame of ERQs at $z\sim2.5$. We then use the {\tt\string CIAO} package to conduct aperture photometry in each of the three bands and in the full 0.3--7\,keV band. We set the aperture radius to be $2''$ and $8''$ for sources observed on the ACIS-S and ACIS-I chips, respectively. The difference is to account for the size increase of the point spread function (PSF) when the targets are imaged with the ACIS-I instead of ACIS-S chips. The background noise level is estimated in an annulus with inner and outer radii being twice and ten times the aperture radius, respectively. 
\textcolor{black}{We consider a source to be detected if the binomial no-source probability \citep{Weisskopf2007} is smaller than 0.0013, which corresponds to $3\sigma$ above the background level in the Gaussian approximation. } 

We use the software package Bayesian Estimation of Hardness Ratios ({\tt\string BEHR}; \citealt{Park2006}) to calculate two hardness ratios defined below:
\begin{equation}\label{eq:hr1}
    \mathrm{HR_1} = \frac{C_\mathrm{m}-C_\mathrm{s}}{C_\mathrm{m}+C_\mathrm{s}}\,\,,
\end{equation}
\begin{equation}\label{eq:hr2}
    \mathrm{HR_2} = \frac{C_\mathrm{h}-C_\mathrm{m}}{C_\mathrm{h}+C_\mathrm{m}}\,\,,
\end{equation}
where $C_\mathrm{s}$, $C_\mathrm{m}$, and $C_\mathrm{h}$ are the photon count in the soft, middle, and hard bands, respectively. {\tt\string BEHR} 
is particularly advantageous in low-count regimes like this study. The code \textcolor{black}{takes the total photon counts in the source and background regions of each energy band,} assumes each photon to be an independent Poisson random variable, and makes posterior draws accordingly irrespective of whether the source is formally detected or not. The non-Gaussian uncertainties are thus propagated correctly. We use a Gaussian quadrature algorithm to calculate the posterior distributions of the HRs with 1000 bins and keep all other parameters as default. This procedure is recommended by the {\tt\string BEHR} package document for doing calculations with less than $\sim15$ counts in either band.

The HRs serve as probes to approximate the level of obscuration along the line of sight between the source and the observer, which we quantify using the hydrogen column density $\nh$. 
We simulate a set of absorbed X-ray spectra using the pre-computed {\tt\string MYTorus} tables \citep{Yaqoob2012}, which simulate the reprocessed X-ray emission from a toroidal structure. \textcolor{black}{Motivated by the viewing geometry inferred from spectropolarimetry observations by \cite{Alexandroff2018}}, we use a {\tt MYTorus} model consisting of a transmitted zeroth-order power-law continuum and a Compton-scattered continuum. We assume a torus inclination angle of $i=85^\circ$, a power-law slope of $\Gamma=1.9$, with the normalization and $N_{\rm H}$ for all components tied together. We assume a range of column densities from $10^{22}\,\cm^{-2}$ to $10^{25}\,\cm^{-2}$ at the source redshift. For completeness, we include an additional Galactic foreground absorption component of $\nh = 10^{20}\,\cm^{-2}$. \textcolor{black}{Using the exact foreground value along each line-of-sight is only produces 1--5\% changes in the final result. This is because Chandra has little sensitivity at $<1.5\,\mathrm{keV}$, which is the regime likely affected by foreground $\nh$. Thus, the foreground virtually has no effect on the results of high-redshift sources.} 

We fold the simulated spectra through the observing-cycle-dependent ACIS-S/ACIS-I response curves. The spectra are then fed into the Portable, Interactive Multi-Mission Simulator ({\tt\string PIMMS}) to simulate count rates as a function of redshift, which we use to calculate the expected $\mathrm{HR_1}$ and $\mathrm{HR_2}$ as a function of redshift $z$ at the different source column densities $\nh$. \textcolor{black}{The time-varying response curve of ACIS-S/ACIS-I cause the same HR values to yield different $\nh$ measurements. Similar to the treatment in \cite{Goulding2018}, we interpolate the $\nh$ values over a grid of redshift and HR$_1$ or HR$_2$ values for each observing cycle involved. We note here that adopting simpler absorbed X-ray spectra such as an absorbed power law rather than the aforementioned {\tt\string MYTorus} models to generate these expected HR values yields only $\sim0.05\,\mathrm{dex}$ difference in the $\nh$ estimation.}
As pointed out by \cite{Goulding2018}, $\mathrm{HR_1}$ is more sensitive to columns below $\sim10^{23}\,\cm^{-2}$, whereas $\mathrm{HR_2}$ is a more reliable probe for columns that are more Compton-thick ($0.3-5\times10^{24}\,\cm^{-2}$). Therefore, similar to their treatment, we adopt the column density inferred from $\mathrm{HR_1}$ if $\mathrm{HR_2}\lesssim-0.75$ (i.e. $\mathrm{HR_2}$ is consistent with no absorption at source redshift) but otherwise we adopt the column density inferred from $\mathrm{HR_2}$. ERQs that are not detected in X-rays have no HR measurements and thus no $\nh$ estimates. \textcolor{black}{The uncertainties on $\nh$ are computed from the $\nh$ values corresponding to the 16th- and 84th-percentile values of the HR posterior distributions.} 

Using {\tt\string PIMMS} and assuming a photon index of $\Gamma=1.9$, we estimate the rest-frame 2--10\,keV absorbed and intrinsic luminosities of our ERQs using their count rates in the observed 0.3--7\,keV band and the inferred $\nh$ values. For sources undetected in the full 0.3--7\,keV band, we compute the $3\sigma$ upper limit of total counts according to \cite{Gehrels1986}, which provides correct treatments on confidence levels for small-number events that follow Poisson statistics. We then use the counts and thus flux upper limits to constrain upper limits for the \textcolor{black}{unabsorbed $L_\mathrm{2-10\,\kev}$ assigning $N_\mathrm{H} = 10^{24}\,\mathrm{cm^{-2}}$} to represent our ignorance of the columns in these X-ray undetected sources -- this choice is also motivated by the possibility that the X-ray-non-detections are due to heavier obscuration. This upper limit also increases with increasing column density that is assumed. \textcolor{black}{We also note that two sources were treated specially. Since SDSS J005044.95$-$021217.6 is only detected in the full 0.3--7\,keV band, no HR and thus $\nh$ can be computed for this source. The HR values obtained from SDSS J134248.86+605641.1 is unable to constrain $\nh$, as the both HR values are below what the HR values that the $\nh$ grid predicts. Thus, we assume $\nh=10^{25}\,\cm^{-2}$ for these two sources and use this value as an upper limit on column density to compute an upper limit on its $L_\mathrm{X}$. }

\section{Results}\label{sec:result}
In this section, we present the results of our joint study of rest-frame UV and optical emission lines and the X-ray properties of the ERQs.

\subsection{ERQ Bolometric Luminosity}
Bolometric luminosity is vital in quasar studies since it is directly related to the rate of accretion that powers these objects. However, it is uncertain to infer bolometric luminosities from X-rays due to the low photon counts and high column densities. Rest-frame optical spectra are not available for all sources in our sample to estimate bolometric luminosity, either. Both the UV emission lines and the UV continuum are likely suppressed by obscuration and reprocessed by scattering in the circumnuclear materials \citep{Hamann2017, Alexandroff2018, ZakamskaAlexandroff2023}. Therefore, we use the mid-infrared (MIR) luminosities to approximate the bolometric luminosities of ERQs \textcolor{black}{in a consistent way since all ERQs have available \textit{WISE} data as part of their selection}. For the easiest comparison with previous works, we measure $\nu L_\nu$ at rest-frame 5 \textcolor{black}{and} 6 microns for the ERQs by fitting a power law on the \textit{WISE} photometry from the AllWISE source catalog \citep{AllWISE}. \textcolor{black}{These luminosity measurements are presented in Table~\ref{tab:sample}.} The MIR light is expected to arise from the isotropic re-emission by the dust surrounding the SMBH and provides the most reliable measure of the bolometric luminosity available to us, although it is possible that light at $6\,\mathrm{\mu m}$ is also suppressed by self-absorption in the torus \citep{Mateos2015}. \cite{Richards2006} find that at 5 or 6 microns, the bolometric correction for type-1 quasars is roughly between 7 and 9, which we adopt to be our estimate. 

\subsection{ERQ Line Properties}\label{sec:erq_line_properties}
\begin{figure*}
    \centering
    \gridline{\fig{j001120_opt.pdf}{0.333\textwidth}{(a)} 
              \fig{j020932_opt.pdf}{0.333\textwidth}{(b)}
              \fig{j093226_opt.pdf}{0.325\textwidth}{(c)}}
    \gridline{\fig{j145623_opt.pdf}{0.6\textwidth}{(d)}
              \fig{j154243_opt.pdf}{0.4\textwidth}{(e)}}
    \gridline{\fig{j080547_civ.pdf}{0.333\textwidth}{(f)}
              \fig{j155057_civ.pdf}{0.333\textwidth}{(g)}
              \fig{j215855_civ.pdf}{0.333\textwidth}{(h)}}
    \caption{ERQs show diverse emission line morphology. \textbf{(a)--(e):} Examples of the multicomponent fit of the \hb+\oiii{} complex and \ha{} complex for the ERQs. The total best-fit model is colored red; the yellow, green, and blue curves represent BLR, NLR, and outflow components, respectively; \feii{} multiplet is displayed in cyan and the continuum is the pink line. \textbf{(f)--(h):} Examples of the multicomponent fit for the \civ$\lambda1549$; solid curves are the total best-fit, and the dashed curves are the different components used (no physical interpretations are associated with these sub-components); shaded regions are masked out during the fitting process.}
    \label{fig:spectra}
\end{figure*}

\begin{figure}
    \centering
    \includegraphics[width=\columnwidth]{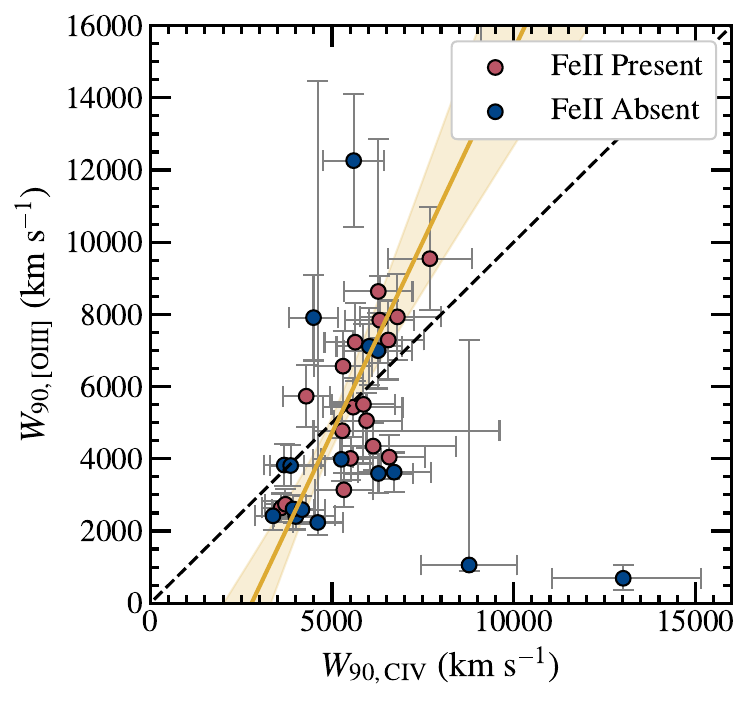}
    \caption{The relationship between the line widths of \oiii$\lambda5008$ and \civ{} of ERQs, sorted by \feii{} presence in the rest-frame optical spectra. The best-fit line is plotted in yellow, and the shaded area marks the $1\sigma$ confidence intervals. The dashed line marks the $1:1$ relation. The point \textcolor{black}{at $W_\mathrm{90,CIV}>10000\,\mathrm{km\,s^{-1}}$ is J1102$-$0075} and is not included in the fit due to poor data quality. }
    \label{fig:civ_oiii_calib}
\end{figure}
\begin{figure}
    \centering
    \includegraphics[width=\columnwidth]{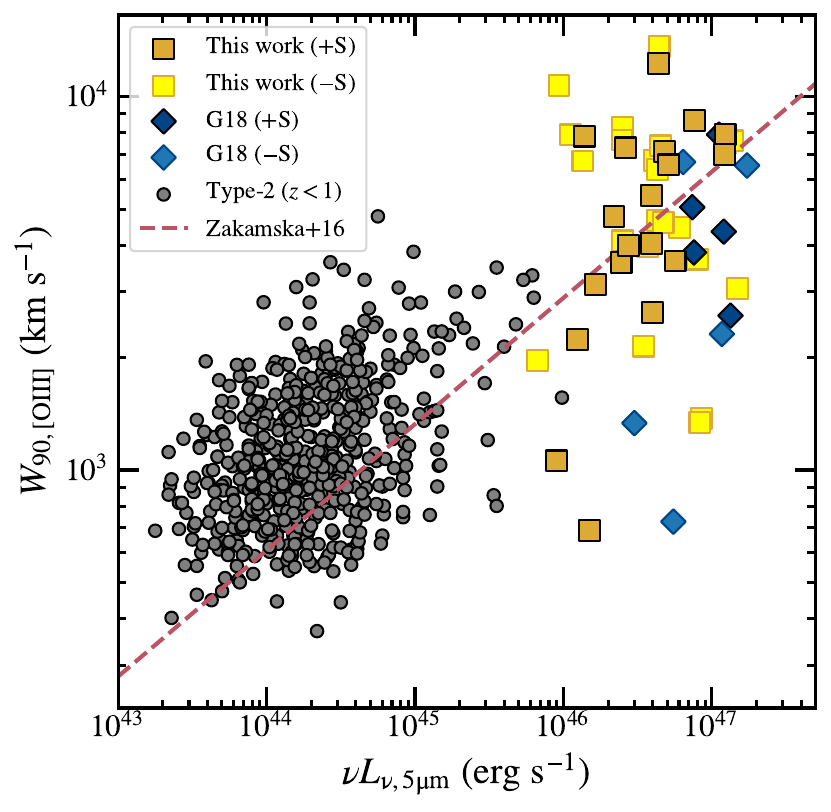}
    \caption{The $W_{90}$ measurements of the \oiii$\lambda5008$ line is plotted as a function of $5\,\mathrm{\mu m}$ luminosity for the ERQs as the yellow circles. The \textcolor{black}{gray circles} are the same measurements for obscured quasars at $z<1$ from \cite{Zakamska&Greene2014}. In the legend, $\pm$S represent whether the sources' $W_\mathrm{90,[OIII]}$ is directly measured from the rest-frame optical spectra ($+$S) or inferred from $W_\mathrm{90,CIV}$ ($-$S). \textcolor{black}{The red dashed line is the relation obtained by \cite{Zakamska2016} for the type-2 quasars and ERQs.}}
    \label{fig:w90_lir} 
\end{figure}

\begin{figure*}
    \centering
    \includegraphics[width=\textwidth]{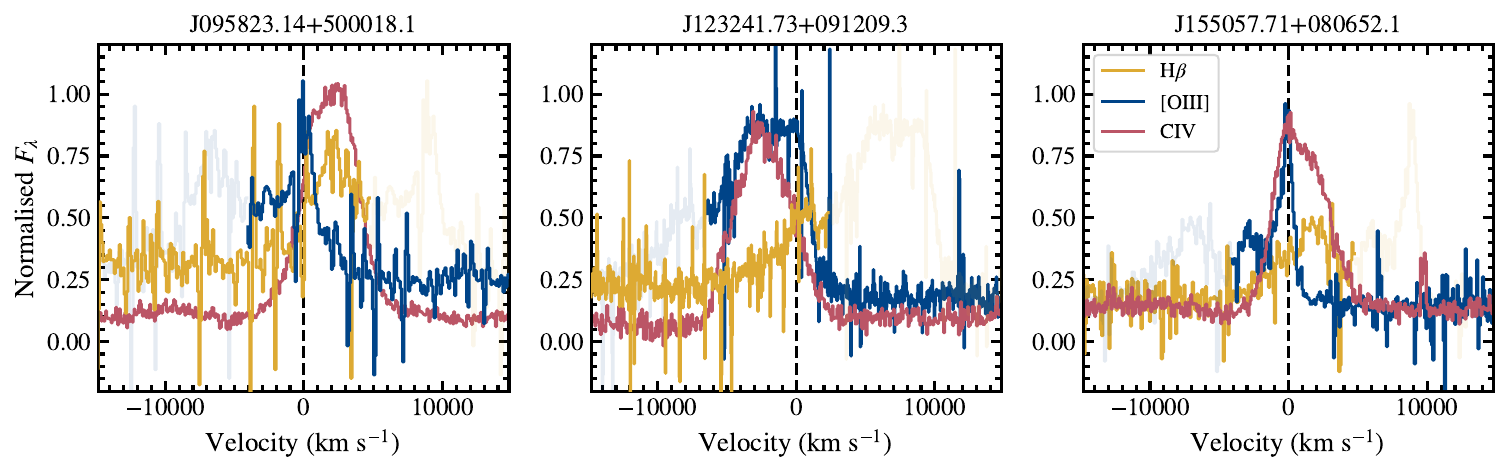}
    \caption{The \hb{}, \oiii{}, and \civ{} emission lines overplotted in velocity space in the rest-frame of \oiii$\lambda5008$ for the three ERQ targets showing signatures of extremely blueshifted NLR (or redshifted BLR) up to $\sim2000\,\mathrm{km\,s^{-1}}$. The faint blue and yellow transparent traces are the full \hb+\oiii{} profiles. }
    \label{fig:strange_spectra}
\end{figure*}
Having settled on the bolometric luminosities, we now present the results from fitting the ERQ spectra. In the rest-frame optical wavelengths, we see a variety of line profiles and spectral features, and a mixture of type-1 and type-2 features is detected (shown in Figure~\ref{fig:spectra}). In some ERQs, a simple NLR+outflow model suffices for fitting all emission lines, but many other ERQs require a broad non-blueshifted component for the Balmer lines. We also detect the permitted \feii{} multiplet pseudo-continuum in 18 sources. We believe the presence of \feii{} and the broad Balmer emissions provide evidence that the BLR in certain ERQs is partially unobscured (see more discussion in \S\,\ref{sec:missing_xray_photons}). We note that SDSS J154243.87+102001.5 cannot be fitted with a quasar+outflow model. Based on the emission-like feature in its fitting residual, additional velocity components such as companions are potentially present.  

In order to compare with previous work, we also measure the line width that encompasses 90\% of the total power, $W_\mathrm{90,line}$, of the \oiii$\lambda5008$ line for ERQs with available rest-frame optical spectra. The $W_\mathrm{90,[OIII]}$ measurements of all but two ERQs are above $2000\,\mathrm{km\,s^{-1}}$. Although we utilize custom fitting routines in this work, we confirm the very broad \oiii{} line widths presented in previous studies of ERQs \citep{Zakamska2016, Perrotta2019}. The two sources with $W_\mathrm{90,[OIII]}<2000\,\mathrm{km\,s^{-1}}$ are SDSS J082649.30+163945.2, which has a low-SNR spectra, and SDSS J110202.68$-$000752.7, whose spectra are affected by strong telluric absorption.

Moreover, in the rest-frame UV, we measure the $W_\mathrm{90,CIV}$ of all ERQs to be above $3000\,\mathrm{km\,s^{-1}}$, consistent with those found in blue type-1 quasars \citep[e.g.,][]{BaskinLaor2005, Sulentic2017}. However, many \civ{} profiles in our ERQ sample also show non-Gaussian and non-Lorentzian shapes. These findings are all consistent with the line properties found in the original ERQ selection paper \citep{Hamann2017} and the physical picture in which \civ{} is partially reprocessed by the outflowing obscuring material \citep[e.g.,][]{Richards2011, Alexandroff2018, Zakamska2003}. We defer any further discussions of the \civ{} physics to future work as it is beyond the scope of this paper. 

\textcolor{black}{Although the \civ{} origin remains ambiguous, a relationship between the line widths may exist if the winds traced by the \civ-emitting gas and \oiii-emitting gas are somehow coupled.} Such a relation can then be used to estimate estimate $W_\mathrm{90,[OIII]}$ for sources without rest-frame optical spectra. Therefore, we fit for a linear $W_\mathrm{90,[OIII]}-W_\mathrm{90,CIV}$ relationship with ERQs with readily available spectra using the Bayesian linear regression code {\tt\string linmix} \citep{Kelly2007} and find
\begin{equation}\label{eq:line_width_calibration}
    W_\mathrm{90,[OIII]} = 2.12_{-0.53}^{+0.64}\times W_\mathrm{90,CIV} - 5923_{-3199}^{+2683}\,\,, 
\end{equation}
where the line widths are in units of $\mathrm{km\,s^{-1}}$ and the errors are the 16th- and 84th-percentiles from the posterior distributions of the slope and the intersect. The best-fit line and the confidence intervals are shown in Figure~\ref{fig:civ_oiii_calib}. The fitted intrinsic scatter for the used line widths is $\sigma_\mathrm{int} = 1138_{-498}^{+549}\,\mathrm{km\,s^{-1}}$. We \textcolor{black}{only} exclude SDSS J110202.68$-$000752.7 from the fit since its \oiii$\lambda5008$ line is overwhelmed by telluric absorption and thus its $W_\mathrm{90,[OIII]}$ cannot be reliably measured. These line widths have a Pearson correlation coefficient of $\rho=0.525$ with a $p$-value of 0.00144. \textcolor{black}{If we were to exclude SDSS J082649.30+163945.2, whose $W_\mathrm{90,[OIII]}$ may be larger than currently measured from its low-SNR spectrum, we would obtain $\rho=0.71$ and $p=4.5\times10^{-6}$.}
Although such calibration may not represent the true \oiii{} line widths, scaling from \civ{} to \oiii{} is the best estimate that we can provide at present given our limited data. All the line width measurements and BLR feature detections are listed in Table~\ref{tab:line_width}.

\textcolor{black}{\cite{Zakamska&Greene2014} find a clear positive correlation between the \oiii$\lambda5008$ line width and 12\,$\mathrm{\mu m}$ luminosities in obscured type-2 quasar at $z<1$ with $\nu L_\mathrm{\nu,12\mu m}\approx 10^{44-46}\,\mathrm{erg\,s^{-1}}$. \cite{Zakamska2016} also find that the line width correlates with $5\,\mathrm{\mu m}$-luminosities as well. The interpretation of such a correlation is that the broadening is due to stronger outflows and winds at higher quasar luminosities.} Following this picture, we show in Figure~\ref{fig:w90_lir} that the ERQs continue to be on this trend, and they are indeed the most wind-dominated and luminous sources in the luminosity-line-width plane. 


\subsection{Kinematically Unassociated Emission Lines in Several ERQs}
We detect ERQs where the Balmer emission, \oiii{}, and/or the \civ{} are significantly offset from each other in velocity space. We show \textcolor{black}{the three sources} in Figure~\ref{fig:strange_spectra}. In SDSS J095823.14+500018.1, both \civ{} and \hb{} are redshifted from \oiii{} by $\sim2000\,\mathrm{km\,s^{-1}}$. The \civ{} in SDSS J123241.73+091209.3 is blueshifted from the other two lines but matches a second peak near \oiii{} that cannot be accounted for by the weaker \oiii$\lambda4960$ alone. In SDSS J155057.71+080652.1, the two kinematic components of \civ{} are each associated with \oiii{} and \hb{}, while the latter two lines themselves are offset from each other by $\sim2000\,\mathrm{km\,s^{-1}}$. This hints at the presence of multiple unresolved sources or that the lines originate from spatially and kinematically different regions in the same source \citep[e.g.,][]{Wylezalek2022, Vayner2023}. More exotic explanations also include SMBH binaries or an off-nucleus recoiling SMBH retaining the BLR but leaving the NLR behind after being ejected in a BH merger event \citep[e.g.,][]{Komossa2012}. We do not see significant \civ{} self-absorption features, which may also alter the emission line profile, in these sources. 

\subsection{ERQ X-ray Properties}\label{sec:erq_xray_properties}

\begin{figure*}
    \centering
    \includegraphics[width=\linewidth]{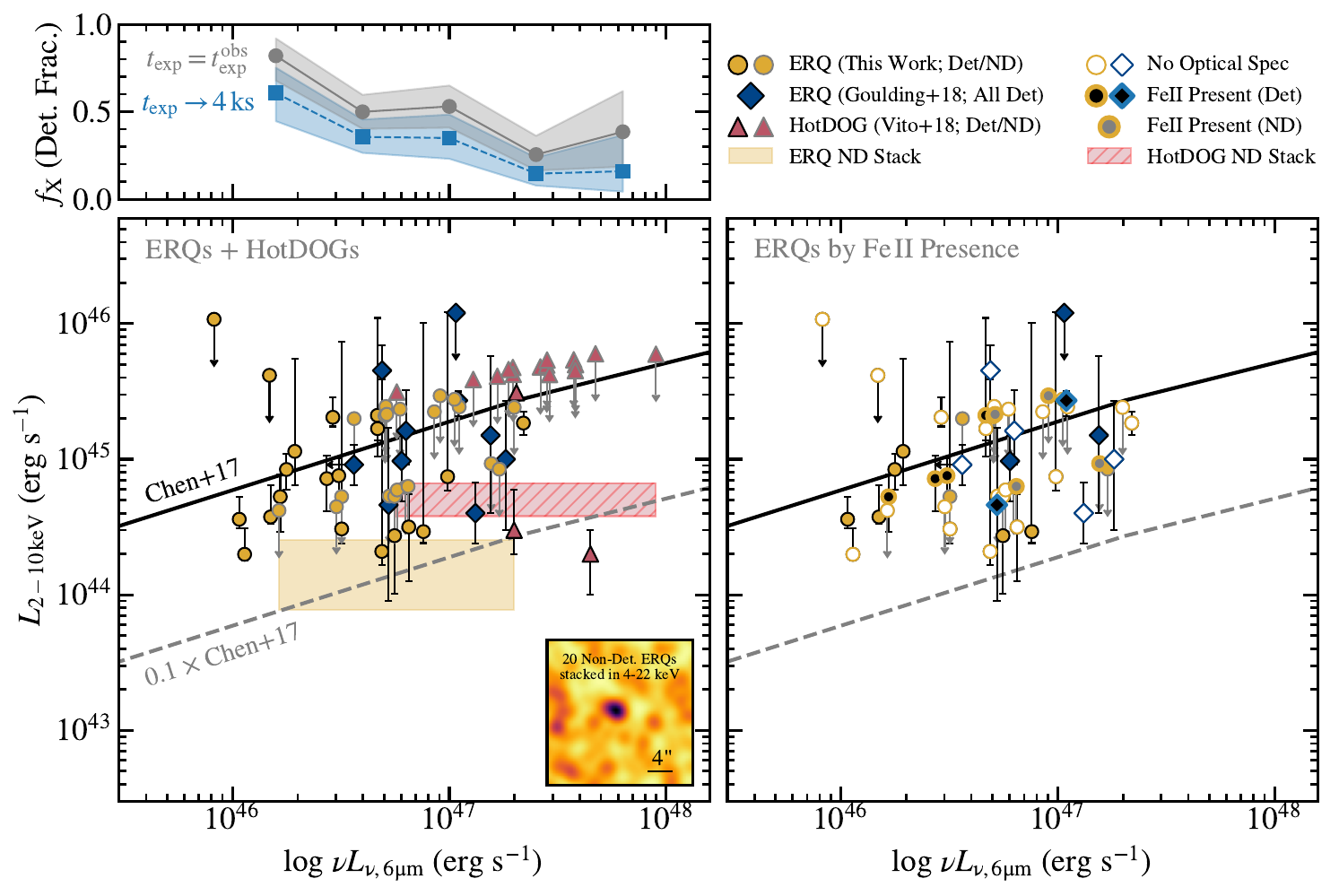}
    \caption{\textbf{Lower Left:} X-ray luminosity is plotted against the mid-infrared luminosity of ERQs and HotDOGs. Empty markers are observed luminosity and filled ones are the absorption-corrected X-ray luminosity. The \textcolor{black}{filled yellow} box represents our luminosity measurements from the X-ray-non-detected ERQ stacks, and the \textcolor{black}{hatched red} box represents the same measurement of \cite{Vito2018} done on the X-ray-non-detected HotDOGs, respectively. The stacked and smoothed image of the X-ray-non-detected ERQs in the rest-frame $4-22$ keV band is shown as the inset, where the circle marks the aperture with which we conduct photometry. 
    \textbf{Lower Right:} we label the ERQs with whether they have rest-frame optical spectra available (empty markers) and whether they have \feii{} lines (black an gray fillings for X-ray-detected and -non-detected sources, respectively) in $L_\mathrm{X}-L_\mathrm{MIR}$ plane. \textcolor{black}{In both panels, the X-ray non-detections are plotted as the $3\sigma$ upper limits of intrinsic $L_\mathrm{X}$ at $N_\mathrm{H} = 10^{24}\,\mathrm{cm^{-2}}$.} \textbf{Top Left:} The solid gray and dashed blue curves track detection fractions of both ERQs and HotDOGs in the observed 0.3--7 keV band as a function of $\nu L_{\nu,6\mu m}$ at the original exposure time and \textcolor{black}{when all scaled down to $\sim4\,$ks}, respectively; the shaded areas are the $1\sigma$ confidence intervals. }
    \label{fig:lx_l6}
\end{figure*}

In Figure \ref{fig:lx_l6}, we show the intrinsic X-ray luminosity $L_\mathrm{2-10\,\kev}$ at rest-frame 2--10$\,\mathrm{keV}$ as a function of mid-infrared luminosity $\nu L_\mathrm{\nu,6\mu m}$ for both the ERQs and HotDOGs. Both populations are included in this figure for their similar SED properties as we outline in \S\,\ref{sec:hotdog_as_comparison_sample}.

In total, 20 ERQ targets in our sample are detected in the full observed 0.3--7$\,\mathrm{keV}$ energy band, and 20 ERQs are not detected. Similarly to previous X-ray studies on ERQs \citep{Goulding2018, Ishikawa2021}, we find that the HR-inferred gas column density of the X-ray-detected ERQs is consistent with sources being heavily obscured. Moreover, $\sim8$ of these ERQs are approaching the Compton-thick regime of $N_\mathrm{H}\geqslant10^{24}\,\mathrm{cm^{-2}}$. Some of these Compton-thick sources are among the sources with the largest number of counts and thus the most luminous targets in our sample. All X-ray measurements on individual targets of our ERQ sample are listed in Table~\ref{tab:xray_measurements}. Of course, as noted already by \cite{Goulding2018}, $\nh$ measurements from HR analysis are far less certain than those derived from X-ray spectra with far more counts. Nevertheless, \textcolor{black}{hardness ratios with appropriate error bars still contain information of the energy spectrum of the sources.} Unfortunately, none of our sources has sufficient counts to produce an X-ray spectrum to improve the $N_\mathrm{H}$ measurements, and only deeper X-ray observations in the future may help resolve this issue.

Visually inspecting the ERQs in the $L_\mathrm{X}-L_\mathrm{MIR}$ space, we find that X-ray-non-detected ERQs are concentrated at \textcolor{black}{$\nu L_\mathrm{\nu,6\mu m} \gtrsim 4\times10^{46}\,\mathrm{erg\,s^{-1}}$.} For the HotDOGs, only 3 out of the 18 targets are detected in 0.3--7$\,\mathrm{keV}$, and the others remain undetected. Given the similarity between ERQs and HotDOGs, we combine the two samples and divide them into 5 equal bins of MIR luminosity.
Within each luminosity bin, we compute the X-ray detection fraction $f_X$ and the associated $1\sigma$ uncertainties by maximizing a binomial likelihood
\begin{equation}
    p(N_\mathrm{det}|f_\mathrm{X},N) = \begin{pmatrix}N\\N_\mathrm{det}\end{pmatrix}f_X^{N_\mathrm{det}}(1-f_X)^{N-N_\mathrm{det}}\,\,,
\end{equation}
where $N$ and $N_\mathrm{det}$ are the number of all sources and X-ray-detected sources, respectively. \textcolor{black}{We find that the X-ray detection fraction overall drops as a function of MIR luminosity (upper panel of Figure \ref{fig:lx_l6}).}

Since our X-ray sample spans a range of exposure times, it is possible that the non-detections may simply be due to shallower exposure times. We thus scale the exposure times of all X-ray observations down to $\sim4\,\mathrm{ks}$ -- the minimum exposure time of our sample -- and run the detection algorithm again. Roughly a third of the X-ray-detected sources become non-detected with this treatment, and we demonstrate in Figure~\ref{fig:lx_l6} that the detection fraction still decreases as a function of MIR luminosity even with the simulated shorter exposure time for each target. \textcolor{black}{Moreover, the mean redshifts of X-ray-non-detected and X-ray-detected ERQs are 2.64 and 2.59, respectively, and the source with the highest redshift ($z\approx3.3$) yield a detection \citep{Goulding2018}. Thus, being higher redshift does not seem to preferentially result in non-detections in our sample.} We conclude that the weakening of X-ray emission as a function of MIR luminosity is physical, as opposed to arising from a selection bias. 

\subsection{Average Properties of X-ray Non-detections}\label{sec:xray_nondet_stack}
To harness the average X-ray properties of the undetected (or weakly detected) sources and to determine if they are intrinsically luminous in X-ray, we stack their \textit{Chandra} images in the rest-frame 4--22 keV band with no weighting assigned to mimic the scenario where we would have observed one individual source over multiple pointings. We repeat the aperture photometry on the stacked image and detect a (stacked) source with total and net counts of \textcolor{black}{19} and \textcolor{black}{15.16}, respectively (see inset of Figure~\ref{fig:lx_l6}). We also use {\tt\string BEHR} to calculate $\mathrm{HR_2}= 0.01\pm0.26$ for the stacked detection. \textcolor{black}{It is not obvious which observing-cycle-dependent $N_\mathrm{H}(z, \mathrm{HR})$ interpolation we should use to compute $N_\mathrm{H}$ from this HR. However, the measured HR value of the stacked sources is consistent with $8\times10^{23}\,\cm^{-2}\lesssim\nh\lesssim3\times10^{24}\,\cm^{-2}$ -- this is 3--9 times denser than the median column density of $\langle N_\mathrm{H}\rangle_\mathrm{det} \approx3\times10^{23}\,\cm^{-2}$ for the X-ray-detected sources. After accounting for the aforementioned range of $N_\mathrm{H}$ values and different observing cycles in {\tt\string PIMMS} modelling and using the median redshift of the X-ray-non-detected ERQs, we estimate the intrinsic X-ray luminosity range to be $10^{43.9}\,\mathrm{erg\,s^{-1}}\lesssim L_\mathrm{2-10\,keV}\lesssim10^{44.4}\,\mathrm{erg\,s^{-1}}$, lower than the unreddened quasars for their $6\,\mathrm{\mu m}$-luminosity range.}


\textcolor{black}{We also present the X-ray luminosity of the stacked X-ray-non-detected HotDOGs by \cite{Vito2018} along with our ERQ stack in Figure~\ref{fig:lx_l6}.} Using HR analysis, we \textcolor{black}{estimate} the X-ray-non-detected HotDOGs to have $N_\mathrm{H}=1.4_{-0.9}^{+0.6}\times10^{24}\,\mathrm{cm^{-2}}$, and the average intrinsic luminosity of the non-detected HotDOGs to be $\log L_\mathrm{2-10\,keV}/\mathrm{erg\,s^{-1}}=44.5_{-0.3}^{+0.1}$ \textcolor{black}{(see red hatched box in Figure 7)} -- both are consistent with \cite{Vito2018}. Similar to the ERQs, the X-ray-non-detected HotDOGs are also heavily obscured and appear underluminous in X-ray \textcolor{black}{for their MIR luminosities}. 

\section{Discussion}\label{sec:discussion}
\textcolor{black}{Using the X-ray observations of a large sample of MIR-selected very luminous quasars hosting extremely powerful outflows, we have found that the ERQ population is overall deficient in X-ray photons for their MIR luminosities. Certain ERQs also show particularly complex kinematic structures based on spectroscopic data. Here, we contextualize our findings, link the X-ray properties to the outflows, and ruminate on the possible implications of our results. }

\subsection{The Missing X-ray Photons}\label{sec:missing_xray_photons}
While many studies have found that unobscured blue quasars show a clear positive correlation between their MIR and X-ray luminosities \citep{Gandhi2009, Stern2015, Chen2017}, a large fraction of the red quasars are X-ray non-detections with the detection fraction dropping at higher MIR luminosity (Figure~\ref{fig:lx_l6}). \cite{Goulding2018} find no X-ray weak ERQs but expect that this population may emerge in surveys with a more extensive search. Indeed, with the sample size now four times larger than the pilot study, we uncover the population of X-ray-weak ERQs. 

One way to make an X-ray-non-detected ERQ is to obscure the source with a Compton-thick X-ray absorbing column. By stacking the X-ray images of the non-detections in the rest-frame 4--22\,keV and making a significant detection at the source position (see \S\,\ref{sec:xray_nondet_stack} and the inset on the left panel of Figure~\ref{fig:lx_l6}), we show that there are indeed X-ray photons leaking out from these sources. This is suggestive of the presence of heavy obscuration at least in some sources with no significant X-ray detection. Such a claim would find support in the X-ray-detected ERQs, which are known to have nearly Compton-thick columns. \textcolor{black}{Even more convincing, the inferred $\nh$ of the non-detection stacks is 3 to 9 times higher than the average $\nh$ of the X-ray-detected population, strongly pointing to the conclusion that the non-detections could be partially attributed to heavier obscurations in these systems.} 
Analogously, both we and \cite{Vito2018} stack the images of the X-ray-non-detected HotDOGs and find that they have comparable gas column density to that of ERQs as well, providing additional circumstantial evidence for the obscuration picture given the similarity between the two populations. 

However, the average intrinsic luminosity of the X-ray-non-detected ERQs is still \textcolor{black}{$\sim$0.5--1$\,\mathrm{dex}$} lower than that expected from the unobscured type-1 quasars of comparable MIR luminosity. Therefore, we suspect that, in addition to being heavily obscured, these non-detected ERQs may be intrinsically X-ray-weak as well. \textcolor{black}{Besides the X-ray stacks, the fact that we find broad-line quasars based on unambiguous detection of \feii{} multiplet emission} in the rest-frame optical provides \textcolor{black}{a potential geometric argument to intrinsic X-ray weakness. Given the \feii{} could be emitted as close to the black hole as $\sim$0.01$-$0.1\,pc \citep{Marinello2016}, their presence in the spectra suggests that we may be looking directly at the BLR and possibly the corona with a less obscured line of sight.} Therefore, the absence of X-ray emission seems more likely to arise from intrinsic weakness than obscuration. The five X-ray-non-detected ERQs with unambiguously detected \feii{} emission are among such examples. Nonetheless, we still cannot rule out geometric effects, where the gas column only obscures the X-ray-emitting corona but not the entire broad-line region. Alternatively, ERQs may have more spatially extended broad-line regions than normal blue quasars. 

\textcolor{black}{In addition to the stacked sources, we} also see some evidence of the combination of heavy obscuration and intrinsic X-ray weakness from the X-ray-detected ERQs. \cite{Goulding2018} find that ERQs show no deviation from the X-ray-to-infrared relations expected for unobscured type-1 quasars. However, with our extended sample, we see the X-ray-detected ERQs on average fall below the type-1 relations found by \cite{Stern2015} and \cite{Chen2017} by a factor as much as $\sim5$ or more (Figure~\ref{fig:lx_l6}). It is not unreasonable to argue that in addition to the measured nearly Compton-thick columns, intrinsic weakness is also responsible for the missing X-ray photons here. As the uncertainties on column densities could be as large as $\sim2\,\mathrm{dex}$ for some targets, it is possible that the luminosities of the X-ray-detected ERQs are spread evenly about the type-1 X-ray-to-infrared relations as well given the current measurement uncertainties. On the other hand, $L_{6 \mathrm{\mu m}}$ may be underestimated for the most obscured sources, making the X-rays even weaker \textcolor{black}{for given MIR luminosity}, relatively speaking.  


\subsection{Linking X-ray Weakness with ERQ Outflows}

\begin{figure}
    \centering
    \includegraphics[width=\columnwidth]{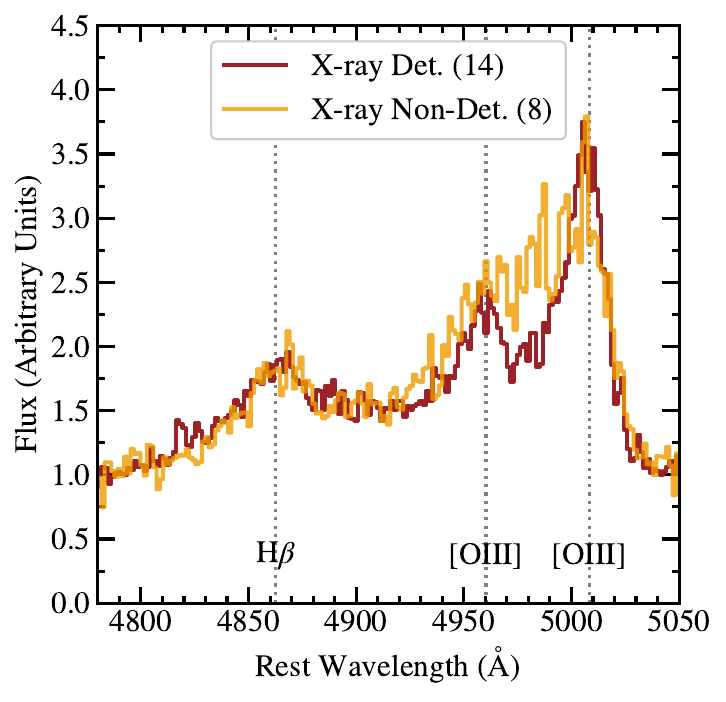}
    \caption{The stacked and normalized \hb+\oiii{} complex for the X-ray-detected and non-detected ERQ targets with rest-frame optical spectra available. J0854+1730 and J1102-0007 are excluded from the stack since their emission line complexes reside close to regions of significant telluric absorption.}
    \label{fig:oiii_stack}
\end{figure}

\begin{figure}
    \centering
    \includegraphics[width=\columnwidth]{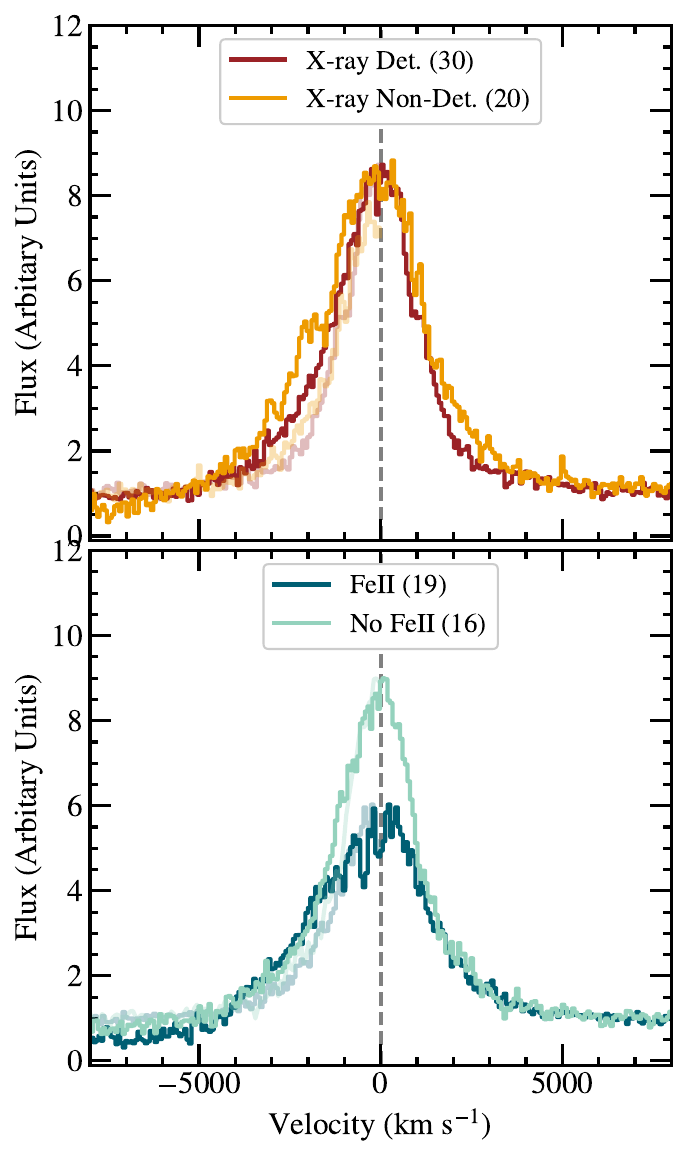}
    \caption{\textbf{Upper Panel:} Stacked and normalized \civ line profiles by X-ray detection. \textbf{Lower Panel:} Stacked and normalized \civ{} profiles by the presence of \feii{} lines in the rest-frame optical spectra. In both panels, the lighter traces at $v<0\,\mathrm{km\,s^{-1}}$ are the reflections of spectra at $v>0\,\mathrm{km\,s^{-1}}$ to show the asymmetry of line profiles. The numbers in the legends are the number of individual spectra stacked in each case.}
    \label{fig:civ_stack}
\end{figure}

\begin{figure}
    \centering
    \includegraphics[width=\columnwidth]{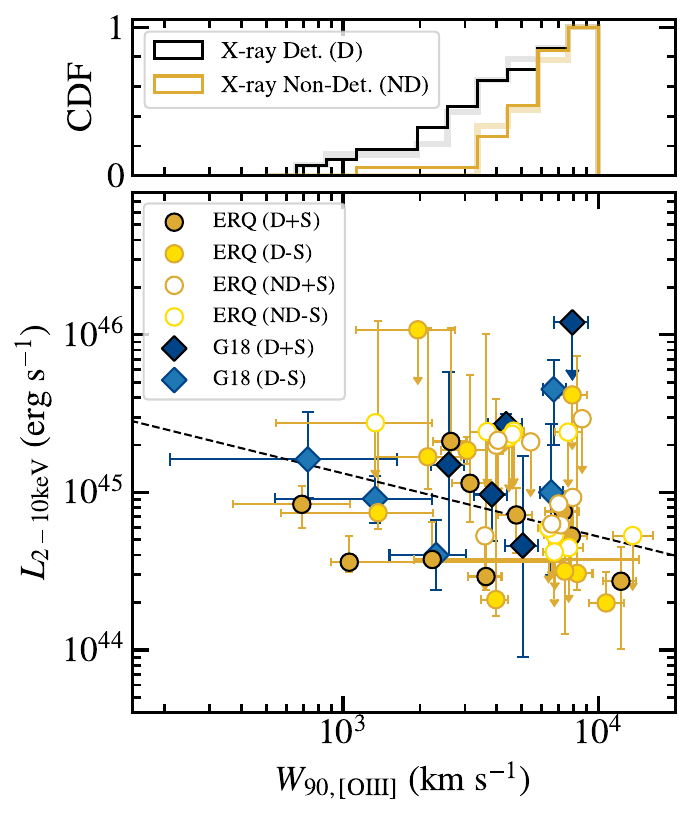}
    \caption{\textbf{Upper Panel:} The more opaque cumulative histogram of \oiii{} line widths for all the X-ray-detected (D) and -non-detected (ND) ERQs. \textcolor{black}{The lighter histograms are produced only using ERQs with rest-frame optical spectra available.} \textbf{Lower Panel}: The intrinsic X-ray luminosity of the ERQs is plotted as a function of \oiii{} line width. }
    \label{fig:lx_w90}
\end{figure}

The natural question to ask next is what physical processes or environments could result in this combination of obscuration and intrinsic X-ray weakness that we observe in ERQs. Luckily, the outflow and winds that are ubiquitously found in this population may help solve the puzzle. Additionally, linking the outflows to the X-ray observations may shed light on the wind-launching mechanisms of the ERQs.

In Figure~\ref{fig:oiii_stack}, we stack the \hb+\oiii{} complex for the X-ray-detected and non-detected ERQs and find that the non-detections show a significantly broader \oiii{} profiles than the X-ray detections \textcolor{black}{($\Delta W_\mathrm{90,[OIII]}\approx 2000-3000\,\mathrm{km\,s^{-1}}$)}. In the stacked \civ{} profiles shown in Figure~\ref{fig:civ_stack}, we see a qualitatively similar difference between the X-ray-detected and non-detected ERQ, although the relative broadening of the non-detections is much less significant than what we see in the \oiii{} profile. This difference between the two emission lines may be due to geometric effects. Since the \civ{} emission could be reprocessed by a thin layer of the dusty torus \citep{Alexandroff2018, ZakamskaAlexandroff2023}, the bulk of the \civ{}-emitting gas may be obscured. This obscuration picture can also be inferred in Figure~\ref{fig:civ_stack}, where the \feii-detected ERQs (less obscured) have a broader \civ{} profile than the ERQs without \feii{} (more obscured). If this is indeed the case, the kinematics of \civ{} reflects only a small portion of the gas covering angle over a small range of distances, whereas the \oiii{} emission is sampled over larger distances and wider covering angles, thus being a better tracer of the quasars' capability to launch winds/outflows. 

\textcolor{black}{Although we have treated the non-detections separately, we do not think X-ray detection divides the ERQs into two physically distinct groups. Rather, the difference in the line widths of the X-ray-detected and -non-detected ERQs could help us infer the wind-launching mechanism of the entire ERQ population.} The correlation between X-ray weakness and strong outflow has always been expected for line-driven winds, a popular mechanism for launching quasar outflows \citep{Schlosman1985, Murray1995}. In large surveys of active galactic nucleus outflows, outflow velocities are overall positively correlated with the bolometric luminosities of the active galactic nuclei \citep[e.g.,][]{Zakamska&Greene2014, Leung2019} -- na\"ively, one would think stronger X-ray emission should be associated with stronger outflow activity. Yet, quasars with the strongest outflow such as BALs \citep[e.g.,][]{Luo2014, Lusso2021, Lusso2023} and some HotDOGs \citep[e.g.,][]{Ricci2017, Vito2018} tend to be associated with intrinsic X-ray weakness. Theoretical work demonstrates that powerful circumnuclear winds can be launched when the gas near the accretion disk absorbs the quasar's UV photons and gets accelerated as a result \citep{Proga1999, Proga2000}. However, this can only occur if the gas is in the proper ionization state to have high bound-bound opacity. An overabundance of X-ray photons at a high enough accretion rate (and thus bolometric luminosity) could overionize the circumnuclear medium, decrease gas opacity, and thus disable the central engine from radiatively launching strong outflows \citep{Proga2000, Proga2004}. 

With this picture in mind, we may explain the observed correlation between X-ray weakness and strong outflow activity. Since \oiii{} is not line-driven \citep[e.g.][]{Dopita2002}, it is plausible that the \oiii-emitting gas is accelerated by the line-driven wind closer to the nuclear region with some yet-to-be-identified coupling mechanisms between the two gas phases. It is also possible that the fast-\oiii{} outflows are driven by radiation pressure on dust \citep{Keating2012, Thompson2015, Ishibashi2018}, which has been claimed to be a common feedback mechanism particularly in luminous red quasar populations \citep{Ishibashi2017}. However, this mechanism does not necessarily require intrinsic X-ray weakness. 

Aside from the emission line stacks (Figure~\ref{fig:oiii_stack} and \ref{fig:civ_stack}), we demonstrate in Figure~\ref{fig:lx_w90} that the X-ray luminosities of the X-ray-detected ERQs may show a negative correlation with \oiii{} line width with a Spearman correlation factor of $r_s=-0.39$ and $p=0.03$. \textcolor{black}{While the p-value is marginal, still this correlation is suggestive, and it is a high priority to confirm with more rest-frame optical spectra.} The cumulative distribution function (CDF) of the \oiii{} line widths for the X-ray non-detections is also preferentially located at higher $W_\mathrm{90,[OIII]}$ values when compared with that for the X-ray-detected ERQs. \textcolor{black}{In Figure ~\ref{fig:lx_w90}, the preference for weaker outflows at higher $L_\mathrm{X}$ is even more obvious in the CDFs produced only using ERQs with directly measured $W_\mathrm{90,[OIII]}$ and excluding those calibrated from \civ{}.} Both panels of Figure~\ref{fig:lx_w90} point towards a potential causal relationship between X-ray weakness and the presence of strong outflow. 

\subsection{Are ERQs Super-Eddington Accretors?}
\textcolor{black}{Given their high bolometric luminosities ($L_\mathrm{bol} \approx 10^{47-48}\,\mathrm{erg\,s^{-1}}$) inferred from MIR luminosities and fast outflows, ERQs are considered to be potential sites of near- or super-Eddington accretions with $L/L_\mathrm{Edd}\approx0.5-10$ at Cosmic Noon \citep{Zakamska2016, Zakamska2019, Perrotta2019}.} Such an assumption may also offer explanations for the observed X-ray weakness in ERQs. At high Eddington ratios, the accretion disk could launch a disk wind into the hot corona. The wind can then Compton-cool the corona by introducing softer photons and thus quenching its ability to up-scatter UV photons into X-ray \citep{Proga2005}. Moreover, it is also possible that the disk wind is so optically thick that it shields the corona and starves it from receiving the seed UV photons to be up-scattered, consequently suppressing the X-ray emission and making an intrinsically X-ray weak quasar. In part, this latter scenario seems to resonate with the high obscuration, and particularly the X-ray-non-detected ERQs. 

Moreover, \cite{Jiang2019} perform global three-dimensional radiation magnetohydrodynamical simulations of super-Eddington accretion onto SMBHs and show that at such high accretion rates, the accretion disk becomes optically thick enough that the accretion power is thermalized due to efficient cooling. If angular momentum in these disks is transported by spiral shocks, sufficient hot gas does not exist to up-scatter the photons in the simulation, consequently making the quasar weak in X-ray. When invoking magnetorotational instability to transport angular momentum in the simulation, hot gas can be produced, but optically thick winds are launched from the super-Eddington disks and obscure the X-ray emitting regions for most observing angles, also resulting in an apparent X-ray weak quasar. However, we note that in this case, the wind is continuum-driven instead of line-driven, as the latter becomes inefficient in the optically thick regime \citep[e.g.,][]{CAK1975, Grafener2017}. In this case, the broader \oiii{} profile may reflect the higher accretion luminosity of these X-ray-non-detected and potentially super-Eddington ERQs; the less significant broadening of \civ{} with respect to X-ray detections may be a result of obscuration and/or the fact that the line-driven mechanism is irrelevant in this scenario. 

\subsection{Complications From the Wider Environment}\label{sec:env_complications}
Up until now, our analysis and discussion assume that each system only has a single black hole in a single galaxy host, but there is the possibility that the high luminosities and complicated line morphologies in both rest-frame optical and UV may be due to multiple objects along the line of sight in some cases. Indeed, galaxy mergers could cause the broad forbidden lines in the ERQs, but \cite{Zakamska2019} find no obvious companions for $\sim8-9$ out of 10 selected ERQ targets based on imaging data from {\it Hubble Space Telescope} ({\it HST}). With \textit{JWST}, \cite{Wylezalek2022} find SDSS J165202.64+172852.3 -- the one source with \textit{HST}-detected companions -- to be a red quasar living in an extremely dense and dynamic protocluster environment, with multiple companion galaxies present on scales of $\sim10\,\mathrm{kpc}$. Thus, in addition to the ubiquitous powerful outflows found in ERQs, unresolved companions and other velocity components in a merger environment may also contribute to the total luminosity and line width of the blended optical emission lines.

The three sources we show in Figure~\ref{fig:strange_spectra} may also support the clustering scenario. For example, the coincidental match-up of the \hb{} and \oiii{} with the two components of \civ{} line in SDSS J155057.71+080652.1 hints at the presence of multiple unresolved companion sources. Although the typical velocity of galaxies in a clustering environment is $600-1000\,\mathrm{km\,s^{-1}}$, projected velocities as high as $\sim2000\,\mathrm{km\,s^{-1}}$ potentially due to clustering has been observed as well \citep{Heckman2009}. Moreover, \cite{Wylezalek2022} point out that the protoclusters may not be virialized yet at $z=2-3$, so producing such line kinematics is not impossible. Even if in fact there is no companion interaction, such observations may also be used to probe the kinematics and structures of different line-emitting regions in the quasar's immediate environment and outflow. The explanations for the significantly kinematically unassociated \civ{}, \oiii{}, and/or \hb{} we mentioned in \S\,\ref{sec:erq_line_properties}, such as recoiling SMBHs or SMBH binaries, all invoke dynamical effects likely introduced by past and/or ongoing merger activity as well. 

Nonetheless, future spatially resolved spectroscopic observations with {\it JWST}, which have already been proven successful in \textcolor{black}{at least one ERQ system \citep{Wylezalek2022}} thanks to the instrument's high angular resolution, would be crucial to studying these ERQ targets. Such observations will help one better constrain the geometry and spatial distribution of the outflowing, X-ray-obscuring material and potential companions in the hosts' immediate environment. With a more extensive survey, one can test whether the protocluster environment would be able to explain the unusually large line widths and other properties of the ERQs \citep{Wylezalek2022} when compared with blue quasars. It could also offer a possibility to test the more exotic hypotheses on the origins of the kinematically unassociated emission lines in certain ERQ targets. 

\section{Conclusion}\label{sec:conclusion}
We present an extensive X-ray study of a total of 50 extremely red quasars at $z=2-3$ selected by their broad \civ$\lambda1549$ line width with proposed \textit{Chandra} observation and archival data. The X-ray data is also supplemented with a spectroscopic sample of 35 ERQs whose rest-frame optical spectra are obtained either from  Magellan/FIRE follow-up observations or by \cite{Perrotta2019}. A total of 24 objects in the spectroscopic sample overlap with the X-ray sources. In this work, we attempt to link the X-ray emission to the powerful outflows ubiquitously found in these quasars by combining ground-based spectroscopic observations with the X-ray data. We summarize the main findings of our analysis as follows. 

To measure the width of the \oiii$\lambda5008$ line and quantify the outflow activity, we perform multi-Gaussian fits of the \hb+\oiii{} complex of the ERQs with available rest-frame optical spectra using a maximum of three components representing the NLR, BLR, and outflows. We find that many ERQs show BLR emission in the rest-frame optical, including the \feii{} multiplet pseudo-continuum and broad and symmetric Balmer emissions. We also conduct multicomponent fits on the \civ$\lambda1549$ emission lines of the ERQs, obtain a line width relation between the \civ{} and \oiii{}, and approximate the \oiii{} line widths for the ERQs without rest-frame optical spectra; \textcolor{black}{nonetheless, it is always the rest-frame optical spectra that would provide the most reliable line width measurements}. All direct \oiii{} line width measurements except for two sources are higher than $2000\,\mathrm{km\,s^{-1}}$, strongly suggesting the presence of powerful outflow. 

On the X-ray end, our analysis reveals a population of previously unknown X-ray-non-detected ERQs. The X-ray detection fraction drops as a function of MIR luminosity. We interpret that such X-ray weakness could be due to heavy obscuration, \textcolor{black}{given that we \textcolor{black}{estimate} the average gas column density of the X-ray-non-detected ERQs to be $\gtrsim10^{24}\,\cm^{-2}$ and higher than that of the X-ray-detected ERQs.} Additionally, we infer that the ERQs also must be intrinsically weak in X-ray since the intrinsic X-ray luminosities of both the detected ERQs (on average) and the stacked non-detections reside below the $L_\mathrm{X}-L_\mathrm{MIR}$ relation for unobscured type-1 quasars. The presence of type-1 spectral features found in the rest-frame optical spectra of the non-detected ERQs may also be indirect evidence of intrinsic X-ray weakness. 

Moreover, we find that the X-ray-non-detected ERQs tend to launch more powerful \oiii-emitting outflows than the X-ray-detected ones do. One option to explain such correlation is that the strong line-driven winds launched at smaller scales, which require intrinsic X-ray weakness, are then coupled with and (potentially mechanically) accelerate the observed \oiii-emitting gas to kpc-scales. Another option is that ERQs are indeed super-Eddington accretors. In this scenario, numerical simulations suggest that such quasars are naturally X-ray weak \citep{Jiang2019}. The larger line widths for the non-detections are due to their higher accretion luminosities, resulting in more powerful continuum-driven winds in the circumnuclear region that subsequently accelerate gas to large scales. 

Lastly, the environments in which ERQs reside may also be complicated. Our spectral fitting reveals that some ERQs show large velocity offset between BLR and NLR emission lines that likely cannot be accounted for with a single SMBH. Protocluster environments, binary SMBHs, or gravitational recoils from SMBH mergers could be invoked to explain such line kinematics. Future spatially resolved spectroscopic observations by \textit{JWST} are required to further study the environmental effect on this quasar population.

\section*{Acknowledgements}
We thank the anonymous referee for the detailed report that leads to improvement of this manuscript. We also acknowledge the support of the grant NSF AAG AWD1007094. This paper employs a list of \textit{Chandra} datasets, obtained by the \textit{Chandra X-ray Observatory}, contained in the Chandra Data Collection 261 (\href{https://doi.org/10.25574/cdc.261}{doi:10.25574/cdc.261}). YM is also grateful for Shengfan Cao's help and suggestions in optimizing the spectral fitting procedures used in this work. 

\startlongtable
\begin{deluxetable*}{cccccccc}
    \centering
    \tablecaption{X-ray Sample of Extremely Red Quasars\label{tab:sample}}
    \tablehead{\colhead{Target} & \colhead{ObsID} & \colhead{RA} & \colhead{Dec} & \colhead{$z$} & \colhead{FWHM(\civ)} & \colhead{$\log \nu L_\mathrm{\nu,6\mu m}$} & \colhead{[O\,{\sc iii}] spec?} \\
    \colhead{} & \colhead{} & \colhead{} & \colhead{} & \colhead{} & \colhead{[$\mathrm{km\,s^{-1}}$]} &  \colhead{[$\,\mathrm{erg\,s^{-1}}$]} & \colhead{} \\
    \colhead{(1)} & \colhead{(2)} & \colhead{(3)} & \colhead{(4)} & \colhead{(5)} & \colhead{(6)} &  \colhead{(7)} & \colhead{(8)}} 
    \startdata
    SDSS J0024$+$2450 & CXO20762 & 00:24:00.25 & $+$24:50:31.9 & 2.7859 & $3523\pm195$ & $47.05\pm0.04$ & $\times$ \\
SDSS J0047$+$2640 & CXO23748 & 00:47:13.21 & $+$26:40:24.7 & 2.5412 & $3685\pm74$ & $46.72\pm0.05$ & $\times$ \\
SDSS J0050$-$0212 & CXO23749 & 00:50:44.95 & $-$02:12:17.6 & 2.2430 & $4343\pm487$ & $46.17\pm0.06$ & $\times$ \\
SDSS J0116$-$0505 & CXO18210 & 01:16:01.43 & $-$05:05:03.9 & 3.1784 & $2291\pm253$ & $47.34\pm0.03$ & $\times$ \\
SDSS J0152$+$3231 & CXO20766 & 01:52:22.58 & $+$32:31:52.7 & 2.7803 & $3677\pm104$ & $46.71\pm0.04$ & $\times$ \\
SDSS J0805$+$4541 & CXO20758 & 08:05:47.66 & $+$45:41:59.0 & 2.2992 & $2667\pm107$ & $46.70\pm0.03$ & \checked \\
SDSS J0826$+$5653 & CXO23750 & 08:26:18.04 & $+$56:53:45.9 & 2.3452 & $3509\pm135$ & $46.22\pm0.06$ & \checked \\
SDSS J0826$+$1639 & CXO23571,24930 & 08:26:49.30 & $+$16:39:45.2 & 2.3111 & $5633\pm150$ & $46.03\pm0.07$ & \checked \\
SDSS J0854$+$1730 & CXO23752 & 08:54:51.11 & $+$17:30:09.1 & 2.6029 & $4199\pm207$ & $46.50\pm0.06$ & \checked \\
SDSS J0958$+$5000 & CXO23753 & 09:58:23.15 & $+$50:00:17.7 & 2.3376 & $4345\pm32$ & $46.75\pm0.03$ & \checked \\
SDSS J1013$+$3427 & CXO20755 & 10:13:24.53 & $+$34:27:02.6 & 2.4582 & $4157\pm51$ & $46.88\pm0.03$ & \checked \\
SDSS J1021$+$2144 & CXO23754 & 10:21:30.74 & $+$21:44:38.4 & 2.1968 & $3567\pm92$ & $46.18\pm0.08$ & \checked \\
SDSS J1025$+$2454 & CXO23755 & 10:25:41.78 & $+$24:54:24.2 & 2.4095 & $5324\pm64$ & $47.19\pm0.01$ & \checked \\
SDSS J1034$+$1430 & CXO23756 & 10:34:56.95 & $+$14:30:12.5 & 2.9560 & $5949\pm367$ & $46.75\pm0.07$ & $\times$ \\
SDSS J1046$+$0243 & CXO20754 & 10:46:11.50 & $+$02:43:51.6 & 2.7702 & $4854\pm126$ & $47.30\pm0.03$ & $\times$ \\
SDSS J1047$+$4844 & CXO20763 & 10:47:18.35 & $+$48:44:33.8 & 2.2747 & $2521\pm51$ & $46.67\pm0.03$ & $\times$ \\
SDSS J1047$+$6213 & CXO23757 & 10:47:54.59 & $+$62:13:00.2 & 2.5398 & $5081\pm133$ & $46.50\pm0.06$ & $\times$ \\
SDSS J1102$-$0007 & CXO23758 & 11:02:02.68 & $-$00:07:52.7 & 2.6174 & $3767\pm282$ & $46.25\pm0.07$ & \checked \\
SDSS J1117$+$4623 & CXO20759 & 11:17:29.56 & $+$46:23:31.2 & 2.1282 & $3053\pm54$ & $46.46\pm0.04$ & $\times$ \\
SDSS J1138$+$4732 & CXO20767 & 11:38:34.68 & $+$47:32:50.0 & 2.3169 & $3296\pm104$ & $46.56\pm0.04$ & \checked \\
SDSS J1217$+$0234 & CXO20756 & 12:17:04.70 & $+$02:34:17.1 & 2.4268 & $2604\pm55$ & $46.67\pm0.04$ & \checked \\
SDSS J1232$+$0912 & CXO23759 & 12:32:41.73 & $+$09:12:09.3 & 2.3906 & $4787\pm52$ & $47.23\pm0.02$ & \checked \\
SDSS J1254$+$2104 & CXO20757 & 12:54:49.50 & $+$21:04:48.4 & 3.1145 & $2482\pm48$ & $46.99\pm0.03$ & $\times$ \\
SDSS J1307$+$3648 & CXO11572 & 13:07:41.38 & $+$36:48:43.0 & 2.3375 & $3633\pm97$ & $46.06\pm0.07$ & $\times$ \\
SDSS J1309$+$5601 & CXO20760 & 13:09:36.14 & $+$56:01:11.3 & 2.5730 & $3630\pm114$ & $46.93\pm0.03$ & $\times$ \\
SDSS J1342$+$6056 & CXO23760 & 13:42:48.86 & $+$60:56:41.1 & 2.4127 & $2652\pm102$ & $45.92\pm0.07$ & $\times$ \\
SDSS J1344$+$4454 & CXO20761 & 13:44:17.34 & $+$44:54:59.4 & 3.0301 & $2871\pm70$ & $47.02\pm0.03$ & $\times$ \\
SDSS J1344$+$1401 & CXO23761 & 13:44:50.51 & $+$14:01:39.2 & 2.7318 & $4487\pm191$ & $46.48\pm0.06$ & $\times$ \\
SDSS J1345$+$6000 & CXO23762 & 13:45:35.66 & $+$60:00:28.4 & 2.9253 & $6460\pm742$ & $46.69\pm0.06$ & $\times$ \\
SDSS J1348$-$0250 & CXO23763 & 13:48:00.13 & $-$02:50:06.4 & 2.2261 & $3654\pm151$ & $46.44\pm0.04$ & \checked \\
SDSS J1355$+$1447 & CXO20765 & 13:55:57.60 & $+$14:47:33.1 & 2.6852 & $2958\pm60$ & $46.77\pm0.03$ & $\times$ \\
SDSS J1456$+$2145 & CXO23764 & 14:56:23.35 & $+$21:45:16.2 & 2.4642 & $4422\pm148$ & $46.49\pm0.03$ & \checked \\
SDSS J1501$+$2317 & CXO23765 & 15:01:17.07 & $+$23:17:30.9 & 3.0169 & $4035\pm130$ & $46.80\pm0.05$ & \checked \\
SDSS J1531$+$1058 & CXO23767 & 15:31:07.14 & $+$10:58:25.8 & 2.7822 & $3937\pm146$ & $46.76\pm0.05$ & $\times$ \\
SDSS J1542$+$1020 & CXO20764 & 15:42:43.87 & $+$10:20:01.5 & 3.2025 & $3901\pm286$ & $46.96\pm0.04$ & \checked \\
SDSS J1550$+$0806 & CXO23766 & 15:50:57.71 & $+$08:06:52.1 & 2.4868 & $4446\pm60$ & $46.29\pm0.07$ & \checked \\
SDSS J1604$+$5633 & CXO20768 & 16:04:31.55 & $+$56:33:54.2 & 2.4925 & $4221\pm82$ & $46.71\pm0.02$ & \checked \\
SDSS J1714$+$4148 & CXO23768 & 17:14:20.38 & $+$41:48:15.7 & 2.3220 & $3816\pm109$ & $46.21\pm0.06$ & $\times$ \\
SDSS J2254$+$2327 & CXO23769 & 22:54:38.30 & $+$23:27:14.5 & 3.0838 & $4412\pm146$ & $46.81\pm0.12$ & $\times$ \\
SDSS J2323$-$0100 & CXO23770 & 23:23:26.17 & $-$01:00:33.1 & 2.3724 & $3989\pm62$ & $46.81\pm0.03$ & \checked \\
\hline\hline
SDSS J0006$+$1215$^\mathrm{G}$ & XMM763780701 & 00:06:10.67 & $+$12:15:01.2 & 2.309 & $3523\pm195$ & $47.16\pm0.01$ (47.04) & \checked \\
SDSS J0220$+$0137$^\mathrm{G}$ & CXO18708 & 02:20:52.13 & $+$01:37:11.4 & 3.318 & $3685\pm74$ & $47.43\pm0.04$ (47.26) & $\times$ \\
SDSS J0826$+$0542$^\mathrm{G}$ & CXO18206 & 08:26:53.42 & $+$05:42:47.3 & 2.578 & $4343\pm487$ & $47.00\pm0.04$ (46.78) & \checked \\
SDSS J0832$+$1615$^\mathrm{G}$ & CXO18207 & 08:32:00.20 & $+$16:15:00.3 & 2.431 & $2291\pm253$ & $47.00\pm0.03$ (46.72) & \checked \\
SDSS J0834$+$0159$^\mathrm{G}$ & XMM762260101 & 08:34:48.48 & $+$01:59:21.1 & 2.591 & $3677\pm104$ & $47.23\pm0.03$ (47.03) & \checked \\
SDSS J0915$+$5613$^\mathrm{G}$ & CXO04821 & 09:15:08.45 & $+$56:13:16.0 & 2.857 & $2667\pm107$ & $46.97\pm0.05$ (46.69) & $\times$ \\
SDSS J1121$+$5705$^\mathrm{G}$ & CXO06958 & 11:21:24.55 & $+$57:05:29.6 & 2.383 & $3509\pm135$ & $46.79\pm0.02$ (46.80) & $\times$ \\
SDSS J1310$+$3225$^\mathrm{G}$ & XMM020540401 & 13:10:47.78 & $+$32:25:18.3 & 3.009 & $5633\pm150$ & $47.15\pm0.02$ (47.12) & $\times$ \\
SDSS J1652$+$1728$^\mathrm{G}$ & CXO18205 & 16:52:02.64 & $+$17:28:52.4 & 2.942 & $4199\pm207$ & $47.22\pm0.02$ (47.19) & \checked \\
SDSS J2129$-$0018$^\mathrm{G}$ & XMM729160501 & 21:29:51.40 & $-$00:18:04.3 & 3.206 & $4345\pm32$ & $<46.54$ ($<46.56$) & $\times$ \\
    \enddata
    \tablecomments{\textcolor{black}{(1) The target name in the SDSS HHMM$\pm$DDMM format. (2) ObsID. (3)--(4) J2000 coordinates. (5) Redshift. (6) Full-width-half-maximum of the \civ$\lambda1549$ emission line. (7) The $6\,\mathrm{\mu m}$-luminosities measured by power-law interpolation on \textit{WISE} photometry; the parenthesized values are measured by \cite{Goulding2018}; on average, the $6\,\mathrm{\mu m}$-luminosity is $\sim1.3$ times larger than the $5\,\mathrm{\mu m}$-luminosity. (8) Does the target have rest-frame optical spectrum available?}}
    \tablecomments{\textcolor{black}{Targets with a superscript ``G" are sources studied by \cite{Goulding2018}.}}
\end{deluxetable*}

\startlongtable
\begin{deluxetable*}{cccccc}
    \centering
    \tablecaption{Line Width Measurements of Extremely Red Quasars\label{tab:line_width}}
    \tablehead{\colhead{Target} & \colhead{$W_\mathrm{90,CIV}$} & \colhead{$W_\mathrm{90,[OIII]]}$} & \colhead{Broad Balmer?} & \colhead{\feii?} & \colhead{Where \oiii{}} \\
    \colhead{} & \colhead{[$\mathrm{km\,s^{-1}}$]} & \colhead{[$\mathrm{km\,s^{-1}}$]} & \colhead{} & \colhead{} & \colhead{} \\
    \colhead{(1)} & \colhead{(2)} & \colhead{(3)} & \colhead{(4)} & \colhead{(5)} & \colhead{(6)}} 
    \startdata
        SDSS J082649.30$+$163945.2 & $8773_{-1316}^{+1316}$ & $1058_{-159}^{+6235}$ & \checked & $\times$ & FIRE \\
        SDSS J085451.11$+$173009.1 & $6278_{-942}^{+942}$ & $3594_{-539}^{+539}$ & $\times$ & $\times$ & FIRE \\
        SDSS J102130.74$+$214438.4 & $4606_{-691}^{+691}$ & $2236_{-335}^{+12242}$ & \checked & $\times$ & FIRE \\
        SDSS J110202.68$-$000752.7 & $13015_{-1952}^{+2148}$ & $689_{-318}^{+372}$ & \checked & $\times$ & FIRE \\
        SDSS J145623.35$+$214516.2 & $6547_{-982}^{+982}$ & $7292_{-1094}^{+1094}$ & \checked & \checked & FIRE \\
        SDSS J150117.07$+$231730.9 & $6026_{-904}^{+904}$ & $7117_{-1067}^{+1067}$ & $\times$ & $\times$ & FIRE \\
        SDSS J154243.87$+$102001.5 & $6272_{-941}^{+941}$ & $8637_{-1296}^{+4211}$ & \checked & \checked & FIRE \\
        \hline\hline
        SDSS J000610.67$+$121501.2 & $6132_{-920}^{+2287}$ & $4352_{-653}^{+653}$ & \checked & \checked & P19 \\
        SDSS J001120.22$+$260109.2 & $5637_{-846}^{+846}$ & $7231_{-1085}^{+1085}$ & \checked & \checked & P19 \\
        SDSS J013413.22$-$023409.7 & $9111_{-1367}^{+1788}$ & $16512_{-2477}^{+2477}$ & \checked & \checked & P19 \\
        SDSS J020932.15$+$312202.7 & $4001_{-600}^{+600}$ & $2403_{-360}^{+360}$ & $\times$ & $\times$ & P19 \\
        SDSS J080547.66$+$454159.0 & $5581_{-837}^{+1358}$ & $5431_{-815}^{+815}$ & \checked & \checked & P19 \\
        SDSS J082618.04$+$565345.9 & $6313_{-947}^{+947}$ & $7841_{-1176}^{+1219}$ & \checked & \checked & P19 \\
        SDSS J082653.42$+$054247.3 & $3684_{-553}^{+553}$ & $3827_{-574}^{+574}$ & $\times$ & $\times$ & P19 \\
        SDSS J083200.20$+$161500.3 & $5948_{-892}^{+976}$ & $5057_{-759}^{+759}$ & \checked & \checked & P19 \\
        SDSS J083448.48$+$015921.1 & $4490_{-673}^{+673}$ & $7911_{-1187}^{+1187}$ & \checked & $\times$ & P19 \\
        SDSS J091303.90$+$234435.2 & $3928_{-589}^{+589}$ & $2611_{-392}^{+392}$ & $\times$ & $\times$ & P19 \\
        SDSS J093226.93$+$461442.8 & $3861_{-579}^{+937}$ & $3814_{-572}^{+572}$ & $\times$ & $\times$ & P19 \\
        SDSS J095823.14$+$500018.1 & $5595_{-839}^{+839}$ & $12262_{-1839}^{+1839}$ & \checked & $\times$ & P19 \\
        SDSS J101324.53$+$342702.6 & $6709_{-1006}^{+1006}$ & $3630_{-544}^{+544}$ & \checked & $\times$ & P19 \\
        SDSS J102541.78$+$245424.2 & $6798_{-1020}^{+1201}$ & $7932_{-1190}^{+1190}$ & \checked & \checked & P19 \\
        SDSS J103146.53$+$290324.1 & $5864_{-880}^{+880}$ & $5508_{-1638}^{+2234}$ & \checked & \checked & P19 \\
        SDSS J113834.68$+$473250.0 & $5255_{-788}^{+788}$ & $3985_{-598}^{+598}$ & \checked & $\times$ & P19 \\
        SDSS J121704.70$+$023417.1 & $3601_{-540}^{+540}$ & $2641_{-396}^{+396}$ & \checked & \checked & P19 \\
        SDSS J123241.73$+$091209.3 & $6262_{-939}^{+939}$ & $6997_{-1050}^{+1050}$ & \checked & $\times$ & P19 \\
        SDSS J134254.45$+$093059.3 & $4289_{-643}^{+643}$ & $5734_{-860}^{+860}$ & \checked & \checked & P19 \\
        SDSS J134800.13$-$025006.4 & $5282_{-792}^{+4328}$ & $4777_{-717}^{+717}$ & \checked & \checked & P19 \\
        SDSS J135608.32$+$073017.2 & $3374_{-506}^{+1717}$ & $2421_{-386}^{+417}$ & \checked & $\times$ & P19 \\
        SDSS J155057.71$+$080652.1 & $5325_{-799}^{+799}$ & $3142_{-471}^{+471}$ & \checked & $\times$ & P19 \\
        SDSS J160431.55$+$563354.2 & $6574_{-986}^{+986}$ & $4046_{-607}^{+607}$ & \checked & \checked & P19 \\
        SDSS J165202.64$+$172852.3 & $4171_{-626}^{+626}$ & $2592_{-389}^{+389}$ & \checked & $\times$ & P19 \\
        SDSS J215855.10$-$014717.9 & $7692_{-1154}^{+1154}$ & $9545_{-1432}^{+1432}$ & \checked & \checked & P19 \\
        SDSS J221524.00$-$005643.8 & $5510_{-826}^{+826}$ & $4010_{-602}^{+602}$ & \checked & \checked & P19 \\
        SDSS J232326.17$-$010033.1 & $5302_{-795}^{+795}$ & $6566_{-985}^{+985}$ & \checked & \checked & P19 \\
        SDSS J232611.97$+$244905.7 & $3717_{-558}^{+571}$ & $2741_{-411}^{+411}$ & \checked & \checked & P19 \\
        \hline\hline
        SDSS J002400.25$+$245031.9 & $4504_{-676}^{+676}$ & $3673_{-536}^{+476}$ & \nodata & \nodata & \civ-calibrated \\
        SDSS J004713.21$+$264024.7 & $5962_{-894}^{+894}$ & $6754_{-647}^{+807}$ & \nodata & \nodata & \civ-calibrated \\
        SDSS J005044.95$-$021217.6 & $6504_{-976}^{+976}$ & $7901_{-885}^{+1118}$ & \nodata & \nodata & \civ-calibrated \\
        SDSS J011601.43$-$050503.9 & $4220_{-633}^{+633}$ & $3060_{-644}^{+576}$ & \nodata & \nodata & \civ-calibrated \\
        SDSS J015222.58$+$323152.7 & $4969_{-745}^{+745}$ & $4662_{-431}^{+428}$ & \nodata & \nodata & \civ-calibrated \\
        SDSS J022052.13$+$013711.4 & $5861_{-879}^{+879}$ & $6540_{-608}^{+754}$ & \nodata & \nodata & \civ-calibrated \\
        SDSS J091508.45$+$561316.0 & $5928_{-889}^{+889}$ & $6682_{-634}^{+789}$ & \nodata & \nodata & \civ-calibrated \\
        SDSS J103456.95$+$143012.5 & $9199_{-1380}^{+1380}$ & $13627_{-2251}^{+2770}$ & \nodata & \nodata & \civ-calibrated \\
        SDSS J104611.50$+$024351.6 & $6369_{-955}^{+955}$ & $7613_{-821}^{+1039}$ & \nodata & \nodata & \civ-calibrated \\
        SDSS J104718.35$+$484433.8 & $3783_{-567}^{+567}$ & $2147_{-819}^{+754}$ & \nodata & \nodata & \civ-calibrated \\
        SDSS J104754.59$+$621300.2 & $6677_{-1002}^{+1002}$ & $8268_{-971}^{+1218}$ & \nodata & \nodata & \civ-calibrated \\
        SDSS J111729.56$+$462331.2 & $4708_{-706}^{+706}$ & $4107_{-475}^{+438}$ & \nodata & \nodata & \civ-calibrated \\
        SDSS J112124.56$+$570529.3 & $2461_{-369}^{+369}$ & $728_{-518}^{+898}$ & \nodata & \nodata & \civ-calibrated \\
        SDSS J125449.50$+$210448.4 & $3342_{-501}^{+501}$ & $1371_{-800}^{+885}$ & \nodata & \nodata & \civ-calibrated \\
        SDSS J130741.38$+$364843.0 & $7835_{-1175}^{+1175}$ & $10727_{-1547}^{+1916}$ & \nodata & \nodata & \civ-calibrated \\
        SDSS J130936.14$+$560111.3 & $4878_{-732}^{+732}$ & $4469_{-440}^{+424}$ & \nodata & \nodata & \civ-calibrated \\
        SDSS J131047.78$+$322518.3 & $3866_{-580}^{+580}$ & $2314_{-794}^{+722}$ & \nodata & \nodata & \civ-calibrated \\
        SDSS J134248.86$+$605641.1 & $3690_{-553}^{+553}$ & $1963_{-835}^{+789}$ & \nodata & \nodata & \civ-calibrated \\
        SDSS J134417.34$+$445459.4 & $3318_{-498}^{+498}$ & $1339_{-794}^{+888}$ & \nodata & \nodata & \civ-calibrated \\
        SDSS J134450.51$+$140139.2 & $6388_{-958}^{+2232}$ & $7655_{-831}^{+1050}$ & \nodata & \nodata & \civ-calibrated \\
        SDSS J134535.66$+$600028.4 & $4642_{-696}^{+696}$ & $3967_{-495}^{+446}$ & \nodata & \nodata & \civ-calibrated \\
        SDSS J135557.60$+$144733.1 & $4941_{-741}^{+741}$ & $4602_{-434}^{+426}$ & \nodata & \nodata & \civ-calibrated \\
        SDSS J153107.14$+$105825.8 & $5788_{-868}^{+868}$ & $6384_{-580}^{+715}$ & \nodata & \nodata & \civ-calibrated \\
        SDSS J171420.38$+$414815.7 & $5948_{-892}^{+892}$ & $6725_{-642}^{+800}$ & \nodata & \nodata & \civ-calibrated \\
        SDSS J212951.40$-$001804.3 & $3316_{-497}^{+497}$ & $1336_{-793}^{+888}$ & \nodata & \nodata & \civ-calibrated \\
        SDSS J225438.30$+$232714.5 & $6260_{-939}^{+939}$ & $7383_{-775}^{+977}$ & \nodata & \nodata & \civ-calibrated \\ 
    \enddata
    \tablecomments{\textcolor{black}{(1) SDSS identifiers. (2)--(3) The line widths encompassing 90\% of the total power of the \civ$\lambda1549$ and \oiii$\lambda5008$ emission lines, respectively. (4) Does the source show broad Balmer emission lines in its rest-frame optical spectrum? (5) Does the source show \feii{} pseudo-continuum? (6) How is the $W_\mathrm{90,[OIII]}$ obtained? ``FIRE" and ``P19" represent direct measurement from our FIRE observation or the rest-frame optical spectra from \cite{Perrotta2019}, respectively, and ``\civ-calibrated" represents that $W_\mathrm{90,[OIII]}$ is inferred from $W_\mathrm{90,CIV}$.}}
\end{deluxetable*}

\startlongtable
\begin{longrotatetable}
\begin{deluxetable*}{cccccccccccc}
    \centering
    \tablecaption{X-ray Measurements of Extremely Red Quasars\label{tab:xray_measurements}}
    \tablehead{\colhead{Target} & \colhead{$t_\mathrm{exp}$} & \colhead{$C_\mathrm{net}^\mathrm{0.3-1}$} & \colhead{$C_\mathrm{net}^\mathrm{1-4}$} & \colhead{$C_\mathrm{net}^\mathrm{4-7}$} & \colhead{$C_\mathrm{net}^\mathrm{0.3-7}$} & \colhead{$\mathrm{HR_1}$} & \colhead{$\mathrm{HR_2}$} & \colhead{HR?} & \colhead{$\log\nh$} & \colhead{$L_\mathrm{X,abs}^\mathrm{2-10\,keV}$} & \colhead{$L_\mathrm{X,int}^\mathrm{2-10\,keV}$}\\
    \colhead{} & \colhead{[ks]} & \colhead{} & \colhead{} & \colhead{} & \colhead{} & \colhead{} & \colhead{} & \colhead{} & \colhead{[$\mathrm{cm^{-2}}$]} & \colhead{[$\mathrm{erg\,s^{-1}}$]} & \colhead{[$\mathrm{erg\,s^{-1}}$]} \\
    \colhead{(1)} & \colhead{(2)} & \colhead{(3)} & \colhead{(4)} & \colhead{(5)} & \colhead{(6)} & \colhead{(7)} & \colhead{(8)} & \colhead{(9)} & \colhead{(10)} & \colhead{(11)} & \colhead{(12)}} 
    \startdata
J0024$+$2450 & 4.1 & $<6.77$ & $<6.85$ & $<6.77$ & $<6.90$ & \nodata & \nodata & \nodata & \nodata & $<44.52$ & $<45.39$ \\
J0047$+$2640 & 20.2 & $<6.95$ & $<7.53$ & $<7.31$ & $<8.24$ & \nodata & \nodata & \nodata & \nodata & $<43.86$ & $<44.73$ \\
J0050$-$0212 & 20.9 & $<6.87$ & $<7.23$ & $<7.11$ & $5.60\pm2.45$ & \nodata & \nodata & \nodata & $<25.0$ & $<43.76$ & $<44.10$ \\
J0116$-$0505 & 70.1 & $<7.46$ & $45.39\pm6.78$ & $25.36\pm5.10$ & $71.46\pm8.54$ & $0.97_{-0.02}^{+0.03}$ & $-0.28_{-0.12}^{+0.11}$ & 2 & $24.1_{-0.1}^{+0.1}$ & $44.30_{-0.08}^{+0.07}$ & $45.27_{-0.09}^{+0.08}$ \\
J0152$+$3231 & 4.1 & $<6.82$ & $<6.79$ & $<6.85$ & $<6.98$ & \nodata & \nodata & \nodata & \nodata & $<44.52$ & $<45.39$ \\
J0805$+$4541 & 4.1 & $<6.79$ & $<7.36$ & $<6.87$ & $<7.53$ & \nodata & \nodata & \nodata & \nodata & $<44.45$ & $<45.32$ \\
J0826$+$5653 & 19.8 & $<6.90$ & $7.71\pm2.83$ & $3.91\pm2.00$ & $11.55\pm3.46$ & $0.96_{-0.06}^{+0.04}$ & $-0.32_{-0.29}^{+0.25}$ & 2 & $23.8_{-0.5}^{+0.2}$ & $44.05_{-0.09}^{+0.09}$ & $44.72_{-0.26}^{+0.16}$ \\
J0826$+$1639 & 22.1 & $2.86\pm1.73$ & $16.65\pm4.12$ & $<7.26$ & $22.30\pm4.80$ & $0.70_{-0.13}^{+0.18}$ & $-0.71_{-0.18}^{+0.13}$ & 2 & $22.7_{-2.7}^{+0.7}$ & $44.43_{-0.08}^{+0.05}$ & $44.56_{-0.06}^{+0.17}$ \\
J0854$+$1730 & 21.1 & $<6.90$ & $<7.55$ & $<7.18$ & $<8.10$ & \nodata & \nodata & \nodata & \nodata & $<43.86$ & $<44.72$ \\
J0958$+$5000 & 17.6 & $<6.85$ & $3.84\pm2.00$ & $<7.05$ & $5.68\pm2.45$ & $0.92_{-0.12}^{+0.08}$ & $-0.34_{-0.42}^{+0.33}$ & 2 & $23.8_{-3.8}^{+0.3}$ & $43.78_{-0.13}^{+0.22}$ & $44.44_{-0.43}^{+0.21}$ \\
J1013$+$3427 & 4.1 & $<6.74$ & $2.96\pm1.73$ & $<6.77$ & $2.95\pm1.73$ & $0.89_{-0.15}^{+0.11}$ & $-0.89_{-0.11}^{+0.15}$ & 1 & $23.0_{-0.7}^{+2.0}$ & $44.27_{-2.79}^{+0.05}$ & $44.47_{-0.08}^{+1.54}$ \\
J1021$+$2144 & 20.8 & $<6.95$ & $22.75\pm4.80$ & $<6.98$ & $25.57\pm5.10$ & $0.92_{-0.05}^{+0.07}$ & $-0.84_{-0.12}^{+0.08}$ & 1 & $22.5_{-2.5}^{+0.9}$ & $44.49_{-0.09}^{+0.03}$ & $44.57_{-0.05}^{+0.24}$ \\
J1025$+$2454 & 9.9 & $<6.87$ & $<7.08$ & $<6.85$ & $<7.31$ & \nodata & \nodata & \nodata & \nodata & $<44.10$ & $<44.97$ \\
J1034$+$1430 & 23.5 & $<6.95$ & $<7.21$ & $<7.16$ & $<7.80$ & \nodata & \nodata & \nodata & \nodata & $<43.86$ & $<44.73$ \\
J1046$+$0243 & 4.1 & $<6.79$ & $<6.85$ & $<6.74$ & $<6.90$ & \nodata & \nodata & \nodata & \nodata & $<44.51$ & $<45.38$ \\
J1047$+$4844 & 4.1 & $<6.79$ & $<6.87$ & $1.99\pm1.41$ & $2.92\pm1.73$ & \nodata & $0.34_{-0.30}^{+0.59}$ & 2 & $24.3_{-0.2}^{+0.7}$ & $43.75_{-2.24}^{+0.21}$ & $45.23_{-0.20}^{+0.81}$ \\
J1047$+$6213 & 24.1 & $<6.90$ & $11.83\pm3.46$ & $<6.98$ & $12.68\pm3.61$ & $0.98_{-0.04}^{+0.02}$ & $-0.85_{-0.15}^{+0.08}$ & 1 & $23.3_{-0.5}^{+1.7}$ & $44.15_{-2.81}^{+0.07}$ & $44.49_{-0.11}^{+1.38}$ \\
J1102$-$0007 & 26.6 & $<6.98$ & $31.62\pm5.66$ & $6.77\pm2.65$ & $38.30\pm6.25$ & $0.99_{-0.01}^{+0.01}$ & $-0.64_{-0.13}^{+0.11}$ & 2 & $23.2_{-3.2}^{+0.3}$ & $44.62_{-0.06}^{+0.14}$ & $44.92_{-0.15}^{+0.11}$ \\
J1117$+$4623 & 4.1 & $5.99\pm2.45$ & $22.94\pm4.80$ & $2.99\pm1.73$ & $31.92\pm5.66$ & $0.58_{-0.13}^{+0.16}$ & $-0.76_{-0.14}^{+0.10}$ & 2 & $22.6_{-2.6}^{+0.6}$ & $45.21_{-0.04}^{+0.03}$ & $45.31_{-0.07}^{+0.15}$ \\
J1138$+$4732 & 4.1 & $<6.90$ & $<6.85$ & $<6.77$ & $<7.03$ & \nodata & \nodata & \nodata & \nodata & $<44.43$ & $<45.30$ \\
J1217$+$0234 & 3.9 & $<6.74$ & $<6.95$ & $1.98\pm1.41$ & $2.90\pm1.73$ & \nodata & $0.36_{-0.26}^{+0.56}$ & 2 & $24.4_{-0.2}^{+0.6}$ & $43.71_{-2.19}^{+0.24}$ & $45.32_{-0.18}^{+0.72}$ \\
J1232$+$0912 & 10.8 & $<6.90$ & $<7.05$ & $<7.00$ & $<7.46$ & \nodata & \nodata & \nodata & \nodata & $<44.06$ & $<44.93$ \\
J1254$+$2104 & 4.1 & $<6.79$ & $3.94\pm2.00$ & $<6.79$ & $3.90\pm2.00$ & $0.92_{-0.11}^{+0.08}$ & $-0.92_{-0.08}^{+0.11}$ & 1 & $23.3_{-0.5}^{+1.7}$ & $44.54_{-2.98}^{+0.09}$ & $44.87_{-0.10}^{+1.21}$ \\
J1307$+$3648 & 39.5 & $<8.77$ & $20.36\pm4.90$ & $<11.67$ & $24.09\pm5.57$ & $0.83_{-0.08}^{+0.14}$ & $-0.88_{-0.12}^{+0.08}$ & 1 & $22.8_{-0.3}^{+0.6}$ & $44.15_{-0.02}^{+0.01}$ & $44.30_{-0.05}^{+0.19}$ \\
J1309$+$5601 & 4.1 & $<6.87$ & $<6.85$ & $<6.79$ & $<7.03$ & \nodata & \nodata & \nodata & \nodata & $<44.48$ & $<45.35$ \\
J1342$+$6056 & 17.1 & $<6.85$ & $10.81\pm3.32$ & $<6.92$ & $12.70\pm3.61$ & $0.82_{-0.09}^{+0.17}$ & $-0.83_{-0.15}^{+0.09}$ & 1 & $<25.0$ & $<41.51$ & $<46.03$ \\
J1344$+$4454 & 4.1 & $<6.77$ & $<6.82$ & $<6.95$ & $<7.05$ & \nodata & \nodata & \nodata & \nodata & $<44.57$ & $<45.44$ \\
J1344$+$1401 & 25.7 & $<6.90$ & $<7.38$ & $<7.21$ & $<7.96$ & \nodata & \nodata & \nodata & \nodata & $<43.78$ & $<44.65$ \\
J1345$+$6000 & 21.8 & $<6.82$ & $6.67\pm2.65$ & $<7.03$ & $6.52\pm2.65$ & $0.95_{-0.07}^{+0.05}$ & $-0.95_{-0.05}^{+0.07}$ & 1 & $23.2_{-1.1}^{+1.8}$ & $44.04_{-2.98}^{+0.13}$ & $44.32_{-0.10}^{+1.27}$ \\
J1348$-$0250 & 13.3 & $<6.87$ & $5.89\pm2.45$ & $3.94\pm2.00$ & $9.77\pm3.16$ & $0.95_{-0.08}^{+0.05}$ & $-0.19_{-0.31}^{+0.28}$ & 2 & $23.9_{-0.4}^{+0.2}$ & $44.08_{-0.11}^{+0.09}$ & $44.86_{-0.24}^{+0.17}$ \\
J1355$+$1447 & 4.1 & $<6.79$ & $<6.85$ & $<6.79$ & $<6.95$ & \nodata & \nodata & \nodata & \nodata & $<44.50$ & $<45.37$ \\
J1456$+$2145 & 21.1 & $2.86\pm1.73$ & $19.75\pm4.47$ & $4.92\pm2.24$ & $27.53\pm5.29$ & $0.74_{-0.11}^{+0.16}$ & $-0.59_{-0.17}^{+0.14}$ & 2 & $23.4_{-3.4}^{+0.3}$ & $44.51_{-0.05}^{+0.15}$ & $44.88_{-0.21}^{+0.13}$ \\
J1501$+$2317 & 20.8 & $<6.87$ & $<7.28$ & $<7.11$ & $<7.75$ & \nodata & \nodata & \nodata & \nodata & $<43.93$ & $<44.79$ \\
J1531$+$1058 & 19.2 & $<6.95$ & $<7.26$ & $<7.03$ & $<7.72$ & \nodata & \nodata & \nodata & \nodata & $<43.90$ & $<44.77$ \\
J1542$+$1020 & 4.1 & $<6.79$ & $<6.87$ & $<6.74$ & $<6.92$ & \nodata & \nodata & \nodata & \nodata & $<44.60$ & $<45.47$ \\
J1550$+$0806 & 17.1 & $<6.87$ & $<7.03$ & $3.92\pm2.00$ & $6.75\pm2.65$ & \nodata & $0.34_{-0.32}^{+0.41}$ & 2 & $24.4_{-0.3}^{+0.6}$ & $43.42_{-2.20}^{+0.32}$ & $45.06_{-0.25}^{+0.68}$ \\
J1604$+$5633 & 4.1 & $<6.77$ & $<6.85$ & $<6.85$ & $<6.98$ & \nodata & \nodata & \nodata & \nodata & $<44.46$ & $<45.33$ \\
J1714$+$4148 & 22.8 & $<7.08$ & $<7.46$ & $<7.08$ & $<8.07$ & \nodata & \nodata & \nodata & \nodata & $<43.75$ & $<44.62$ \\
J2254$+$2327 & 25.6 & $<6.95$ & $3.68\pm2.00$ & $<7.33$ & $5.35\pm2.45$ & $0.91_{-0.12}^{+0.09}$ & $-0.35_{-0.44}^{+0.34}$ & 2 & $23.9_{-3.9}^{+0.3}$ & $43.72_{-0.27}^{+0.37}$ & $44.50_{-0.40}^{+0.24}$ \\
J2323$-$0100 & 14.9 & $<6.98$ & $<7.18$ & $<7.08$ & $<7.72$ & \nodata & \nodata & \nodata & \nodata & $<43.93$ & $<44.80$ \\
\enddata
\tablecomments{\textcolor{black}{(1) The target name in the SDSS HHMM$\pm$DDMM format. (2) Exposure times. (3)--(5) Net counts within the observed 0.3--1\,keV, 1--4\,keV, and 4--7\,keV bands. (6)--(7) Hardness ratios defined by Equations~\ref{eq:hr1} and \ref{eq:hr2}. (8) Which HR is used to estimate gas column density? (9) Gas column densities. (10) X-ray luminosities within rest-frame 2--10\,keV before correcting for gas absorption. (11) Intrinsic X-ray luminosities within rest-frame 2--10\,keV after gas absorption correction.}}
\tablecomments{\textcolor{black}{{\tt\string BEHR} is able to compute HRs even if one of the two bands is undetected. Hence, HR values are not presented only for sources that are undetected in both relevant energy bands.}}
\end{deluxetable*}
\end{longrotatetable}

\bibliography{ref}{}
\bibliographystyle{aasjournal}

\end{CJK*}
\end{document}